\documentclass[showpacs,aps,prd,nofootinbib,floatfix,amsmath,amssymb]{revtex4}
\usepackage{graphicx}
\usepackage{slashed}
\usepackage{mathtools}
\usepackage{multirow,array}
\usepackage{tikz}
\usetikzlibrary{decorations.markings}
\begin{document}

\makeatletter
%Feynman slash
\newbox\slashbox \setbox\slashbox=\hbox{$/$}
\newbox\Slashbox \setbox\Slashbox=\hbox{\large$/$}
\def\pFMslash#1{\setbox\@tempboxa=\hbox{$#1$}
  \@tempdima=0.5\wd\slashbox \advance\@tempdima 0.5\wd\@tempboxa
  \copy\slashbox \kern-\@tempdima \box\@tempboxa}
\def\pFMSlash#1{\setbox\@tempboxa=\hbox{$#1$}
  \@tempdima=0.5\wd\Slashbox \advance\@tempdima 0.5\wd\@tempboxa
  \copy\Slashbox \kern-\@tempdima \box\@tempboxa}
\def\FMslash{\protect\pFMslash}
\def\FMSlash{\protect\pFMSlash}
\def\miss#1{\ifmmode{/\mkern-11mu #1}\else{${/\mkern-11mu #1}$}\fi}
%%%% Uso:  \pFMSlash{p}
\makeatother

%\tightenlines
\title{One-loop structure of the photon propagator in the Standard Model Extension}
\author{A. I. Hern\' andez-Ju\' arez$ {}^{(a)}$}
\author{J. Monta\~no$ {}^{(b,c)}$}
\author{ H. Novales-S\' anchez$ {}^{(a)}$}
\author{ M. Salinas$ {}^{(a)}$}
\author{J. J. Toscano${}^{(a)}$}
\author{ O. V\' azquez-Hern\' andez${}^{(a)}$}
\affiliation{$^{(a)}$Facultad de Ciencias F\'{\i}sico Matem\'aticas,
Benem\'erita Universidad Aut\'onoma de Puebla, Apartado Postal
1152, Puebla, Puebla, M\'exico.\\
$^{(b)}$ Facultad de Ciencias F\'\i sico Matem\' aticas, Universidad Michoacana de San Nicol\' as de Hidalgo,
Avenida Francisco J. M\' ujica S/N, 58060, Morelia, Michoac\' an, M\' exico.\\
$^{(c)}$ CONACYT, M\' exico.
}
\begin{abstract}
Radiative corrections on the photon propagator from the electroweak sector are studied in the context of the minimal Lorentz- and $CPT$-violating Standard Model Extension, with focus on the Yukawa, Higgs and gauge sectors. The most general Lorentz-violating ghost sector dictated by BRST symmetry and renormalization theory is derived. We stress the introduction of a Lorentz-violating nonlinear gauge that greatly simplifies both the Higgs-sector extension and the gauge-sector extension, which can be very helpful in radiative corrections. At one loop, these sectors contribute to the {\it CPT}-even part of the photon propagator, which is characterized by the Riemann-type tensor $(k_F)_{\alpha \beta \mu \nu}$. Exact results for the contributions to the ${\rm SO}(1,3)$ irreducible parts of $(k_F)_{\alpha \beta \mu \nu}$, namely, the Weyl-type tensor $(\hat{k}_F)_{\alpha \beta \mu \nu}$, the Ricci-type tensor $(k_F)_{\alpha \beta}$, and the curvature-type scalar $k_F$, are presented. In the Yukawa sector, with general flavor-violating effects, all the one-loop contributions are ultraviolet finite, but most of them are unobservable due to finite renormalization of the field, the electric charge, $(\hat{k}_F)_{\alpha \beta \mu \nu}$, and $(k_F)_{\alpha \beta}$. The only observable effect is a contribution proportional to $(k_F)_{\alpha \beta}$ that emerges via a dimension-6 term that is both observer and gauge invariant. In the Higgs and gauge sectors, all the irreducible parts of the corresponding Riemann-type tensors receive divergent contributions, so they are observable. The only finite contribution corresponds to the previously-mentioned dimension-6 term. By thinking of these contributions as a radiative correction to the renormalized tensors, and assuming that both effects are of the same order of magnitude, bounds from vacuum birefringence are derived and compared with results in the literature. Bounds on contributions proportional to $(k_F)_{\alpha \beta}$, innocuous to birefringence, are also derived using limits imposed on the renormalized tensor from the Laser-Interferometer-Gravitational-Wave-Observatory data. We compare these bounds with those already existing in the literature. The beta functions associated with the  $(\hat{k}_F)_{\alpha \beta \mu \nu}$ and $(k_F)_{\alpha \beta}$ tensors are derived.
\end{abstract}


\maketitle

\section{Introduction}
\label{I}There are well-founded reasons to suspect that Lorentz invariance is not strictly maintained at the Planck scale. At a fundamental level, clues of Lorentz violation (LV) arise from efforts to merge quantum theory and general relativity into a unified theory. While it is true that local Lorentz invariance is a central feature of general relativity, consistence with quantum theory may require some kind of modification in its structure, suggesting the existence of an underlying preferred time that would radically change our notion of time~\cite{TP}. Explicit or spontaneous LV has been studied in the context of string theory~\cite{ST}, noncommutative geometry~\cite{NC}, loop quantum gravity~\cite{LQG}, cosmologically varying scalar~\cite{CVS}, random-dynamics models~\cite{RDM}, Horava-Lifshitz theories~\cite{HTT}, and brane-world scenarios~\cite{BW}. At low energies, effects of Lorentz and {\it CPT} violation can be described in a model independent way by the so-called {\it Standard Model Extension} (SME), which is an effective field theory that contains General Relativity and the Standard Model (SM)~\cite{GRSME,SME}. In its minimal version (mSME)~\cite{SME}, the model contains only renormalizable interactions, in the sense of mass units, but non-renormalizable interactions are expected to play a dominant role at higher energies~\cite{NRO}. Implications of non-renormalizable interactions have been studied in diverse contexts, such as electrodynamics with operators of arbitrary dimension~\cite{NRSME1}, quantum-field theoretic properties~\cite{NRSME2}, aspects of electromagnetic properties of spin-1/2 particles~\cite{NRSME3}, some phenomenological implications in the electroweak sector~\cite{NRSME4}, and lepton-flavor non-conservation~\cite{LFVSME}.

In this paper, we are interested in studying one-loop LV effects on the photon propagator in the context of the mSME. One-loop effects from LV electromagnetic currents have already been addressed in the context of the QED extension(QEDE)~\cite{QEDE}. In this work, we study the corresponding  radiative effects from the Yukawa, Higgs and gauge sectors of the mSME. The renormalizable pure-photon sector of the QEDE is given by the following Lagrangian~\cite{SME}:
\begin{equation}
\label{PL}
{\cal L}_{\rm photon}=-\frac{1}{4}F_{\mu \nu}F^{\mu \nu}-\frac{1}{4}(k_F)_{\lambda \rho \mu \nu}F^{\lambda \rho}F^{\mu \nu}+\frac{1}{2}(k_{AF})^\rho \epsilon_{\rho \lambda \mu \nu}A^\lambda F^{\mu \nu}\, .
\end{equation}
Even though the action $\int d^4x\,{\cal L}_{\rm photon}$ is invariant under both gauge transformations and observer spacetime-coordinates transformations, this Lagrangian is gauge invariant only up to a total derivative due to the presence of the Chern-Simons-like (CS) term, distinguished by the coefficient $k_{AF}$. The CS term, which characterizes the {\it CPT}-odd Carroll-Field-Jackiw electrodynamics~\cite{CFJ}, has very interesting properties that have been widely studied in the literature in various contexts~\cite{CG,JK,PVV,VRC,SYMM,CF, Ind1,Ind2,Ind3,Ind4,Ind5,Ind6}.  Photon birefringence in vacuum is one important implication of the CS term~\cite{SME,JBi}; non-observation of such a phenomenon imposes a very stringent bound, $\sim10^{-42}\, {\rm GeV}$, on its components~\cite{BiB}. In the mSME, the CS term can be radiatively induced~\cite{JK,PVV} via the {\it CPT}-odd term $b_\mu \bar{\psi} \gamma^\mu \gamma^5\psi$ of the fermion sector, with $b_\mu$ a constant 4-vector. The presence of this term leads to the relation $(k_{AF})_\mu=c\,b_{\mu}$, where $c$ is a finite one-loop amplitude. However, since the amplitude $c$ is regularization-scheme dependent and is thus undetermined~\cite{Ind1,Ind2,Ind3,Ind4,Ind5,Ind6}, the stringent bound on $k_{AF}$ cannot be automatically implemented on $b$.

On the other hand, the $k_F$ term, which characterizes the {\it CPT}-even Lorentz-violating part of Eq.~(\ref{PL}), has also been the subject of considerable interest in the literature. Its presence in the classical action is required by renormalization theory~\cite{QEDE} because, in contrast with the {\it CPT}-odd $k_{AF}$ term, it receives divergent contributions at one loop~\cite{SME,QEDE}. The tensor $(k_F)_{\lambda \rho \mu \nu}$ has the same properties as the Riemann tensor, so, in principle, it has 20 independent components, some of which are restricted by photon birefringence. In order to classify the birefringent and the non-birefringent parts of $(k_F)_{\lambda \rho \mu \nu}$ on the grounds symmetry criteria, and for the sake of our own purposes, it is convenient to work with the ${\rm SO}(1,3)$ irreducible parts of this tensor. In four dimensions $(k_F)_{\lambda \rho \mu \nu}$ can be decomposed into its irreducible parts as
\begin{equation}
\label{RT1}
(k_F)_{\lambda \rho \mu \nu}=(\hat{k}_F)_{\lambda \rho \mu \nu}+(\tilde{k}_F)_{\lambda \rho \mu \nu}+\frac{k_F}{6}\left(g_{\lambda \mu}g_{\rho\nu}-g_{\rho \mu}g_{\lambda \nu}\right)\, ,
\end{equation}
where
\begin{equation}
\label{RT2}
(\tilde{k}_F)_{\lambda \rho \mu \nu}=\frac{1}{2}\left[g_{\rho \mu}(k_F)_{\lambda \nu} - g_{\rho \nu}(k_F)_{\lambda \mu}+g_{\lambda \nu}(k_F)_{\rho \mu}      -g_{\lambda \mu}(k_F)_{\rho \nu}\right]\, .
\end{equation}
In the above expressions we have the following tensors: $(\hat{k}_F)_{\lambda \rho \mu \nu}$ \footnote{This tensor is analogous of the Weyl tensor, in the context of general relativity.}, which has the same symmetries as $(k_F)_{\lambda \rho \mu \nu}$, is defined so that every tensor contraction between indices equals 0 and thus has ten independent components; $(k_F)_{\rho \nu}=g^{\lambda \mu}(k_F)_{\lambda \rho \mu \nu}$ is a symmetric tensor (analogous of the Ricci tensor); and $k_F=g^{\mu \nu}(k_F)_{\mu \nu}$ is a scalar (analogous of the scalar curvature). Notice that if we assume $k_F=0$, then $(k_F)_{\rho \nu}$ becomes a traceless symmetric tensor with nine components, which in turn implies that $(k_F)_{\lambda \rho \mu \nu}$ is a doubly-traceless tensor. This is usually assumed in the literature because the trace can be absorbed by a redefinition of the electromagnetic field. In what follows we assume, for clarity purposes, that  $k_F\neq 0$, but always keep in mind our previous comment.

An important point worth emphasizing about the decomposition of $k_{F}$ into its irreducible parts is that $\hat{k}_F$, usually parametrized in the literature by the $3\times 3$ matrices $\tilde{\kappa}_{e^+}$ and $\tilde{\kappa}_{o^-}$, is sensitive to birefringence~\cite{SME,BKF,Hee}, whereas $\tilde{k}_F$ is not~\cite{Casana,Brett}. In the literature, the non-birefringent tensor $(\tilde{k}_F)_{\mu \alpha \beta \nu}$ is typically parametrized in terms of the $3\times 3$ matrices $\tilde{\kappa}_{e^-}$ and $\tilde{\kappa}_{o^+}$, and the parameter $\tilde{\kappa}_{tr}$. The components $\hat{k}_F$ would be restricted to be less than $10^{-32}$~\cite{BKF}. Some sectors of the mSME, such as those that involve charged fermions or the $W$ gauge boson, can contribute to $k_{F}$ at one loop. Due to stringent constraints from birefringence, it is important to know, from a phenomenological point of view, which sectors of the mSME can contribute to  $\hat{k}_F$ and which ones cannot. Several scenarios may occur. ($a$) there are contributions to all the irreducible parts of  $k_{F}$ but they are free of ultraviolet divergences. In this scenario, such contributions do not translate into limits on the sector under consideration because they can be absorbed in redefinitions of the field and parameters of the Lagrangian given in Eq.~(\ref{PL}), being therefore unobservable. As we show in the present work, this is the case of the {\it CPT}-even extensions of the Lepton and Quark Yukawa sectors. The fact that these sectors elude (at least at the one-loop level) the strong constraints imposed by birefringence is important because they induce flavor violation at tree level, which is a subject of current interest. To our knowledge, effects from the most general extension of the Yukawa sector on the photon propagator have not been considered so far. Bounds on the first families of leptons and quarks, respectively $\sim10^{-18}$ and $\sim10^{-21}$, have been derived from penning-trap electron-positron and proton-antiproton  experiments~\cite{He}. Regarding the Lepton sector, bounds on flavor-violating transitions $\mu-e$ and $\tau-\mu$ of orders $10^{-34}$ and $10^{-21}$ have been respectively derived from experimental constraints on the decays $\mu \to e\gamma$ and $\tau \to \mu \gamma$~\cite{LFVSME}. ($b$) there are contributions to all irreducible parts  of $k_F$ and they are divergent. In this scenario, severe restrictions on some parts of the sector under consideration can be imposed from birefringence once a renormalization scheme has been implemented~\cite{SME,QEDE}. This occurs in certain class of contributions that emerge from the extensions of both the Higgs sector and the gauge electroweak sector. In this case, there are pieces that contribute to both irreducible parts $(\hat{k}_F)_{\mu \alpha \beta \nu}$ and $(\tilde{k}_F)_{\mu \alpha \beta \nu}$ of the tensor $(k_F)_{\mu \alpha \beta \nu}$, but there are other interactions that only contribute to the non-birefringent part  $(\tilde{k}_F)_{\mu \alpha \beta \nu}$. In Ref.~\cite{Sher}, bounds of order $\sim10^{-16}$ for those  contributions to $(\hat{k}_F)_{\mu \alpha \beta \nu}$ were derived from birefringence. In the same reference, the authors used a coordinate and field transformation to relate the non-birefringent contributions to the fermion sector and used previous limits on this sector to bound it. However, as argued in~\cite{OBS}, this methodology for indirectly deriving bounds is doubtful, so the limits obtained in~\cite{Sher} are not reliable. In the present paper, we use recent bounds~\cite{BNBR} on the parameters $\tilde{\kappa}_{e^-}$ and $\tilde{\kappa}_{o^+}$, which were derived from preliminary data obtained by the Laser Interferometer Gravitational-Wave Observatory, to directly constrain these non-birefringent pieces of the Higgs sector.

 Radiative effects from the Higgs sector must be treated carefully because this scalar sector is connected to the gauge sector through the Higgs mechanism. This observation is even more relevant when considering radiative corrections on off-shell Green's functions, which is the case addressed in the present work. The quantization of a gauge sector and a Higgs sector linked by the Higgs mechanism cannot be carried out separately. This means that Higgs effects naturally arise in the corresponding ghost sector. Depending on the gauge-fixing procedure used to quantize the theory, one can introduce strong modifications, not only in the original Higgs sector, but also in the ghost sector and in the Yang-Mills sector. A judicious choice of gauge-fixing functions can greatly simplify loop calculations. Motivated by the high level of complexity of loop calculations within the SME, we explore the possibility of introducing a gauge-fixing procedure that aims at simplifying the calculation of radiative corrections as much as possible. Based on the Becchi-Rouet-Stora-Tyutin (BRST)~\cite{BRST1,BRST2,BRST3} symmetry and on renormalization theory as well, we propose nonlinear gauge-fixing functions that depend not only on SM physical parameters but also on the constant Lorentz tensors that appear in the SME. Our gauge-fixing procedure extension is a natural generalization of nonlinear gauge-fixing procedures introduced in the SM~\cite{NLG} and in other contexts~\cite{NLGG}. However, we would like to emphasize that the introduction of a Lorentz-violating gauge-fixing procedure obeys to practical purposes rather than to some fundamental motivation, since the SME can be quantized in exactly the same way as the SM. The reason for this is that both the SM and SME are governed by the same gauge-symmetry group and no new degrees of freedom are introduced.

 Another goal of this work is the calculation of the beta function associated with the tensor $(k_F)_{\lambda \rho \mu \nu}$. The contribution to this beta function from LV electromagnetic currents was already calculated~\cite{QEDE} in the context of QEDE. In this work, we include the contributions from the extensions of the Higgs and gauge sectors and show that there is no contribution from the Yukawa-sector extension.

The rest of the paper has been organized as follows. In Sec.~\ref{FW}, we present a brief discussion on the Yukawa, Higgs, and gauge sectors of the mSME. Sec.~\ref{GFP} is devoted to discuss the most general gauge-fixing procedure for the SME, allowed by BRST symmetry and renormalization theory. In Sec.~\ref{FPP}, the one-loop contribution from the Yukawa sector to the photon propagator is presented. Sec.~\ref{HS} is dedicated to discuss the one-loop contribution from the Higgs sector to the photon propagator. Sec. \ref{GS} is devoted to calculate the one-loop contribution from the gauge-sector extension. In Sec.~\ref{bF}, the one-loop structure of the beta function for the $(k_F)_{\lambda \rho \mu \nu}$ tensor is discussed in the context of the mSME. In Sec.~\ref{C}, a summary of our results is presented. A brief review of the field-antifield formalism is presented in the context of the SM gauge group in Appendix~\ref{FAFF}; while form factors appearing in the amplitudes are listed in Appendices \ref{FF} and \ref{BF}.

\section{Framework}
\label{FW}
The main goal of the present work is the investigation of the impact of one-loop radiative corrections on the tensor $(k_F)_{\lambda \rho \mu \nu}$, defined in the Lagrangian given in Eq.~(\ref{PL}), in the context of the mSME. A complete calculation of the contributions induced by the Yukawa, Higgs, and gauge sectors will be presented. We now proceed to discuss these sectors of the mSME.

\subsection{The Yukawa-sector extension}

In the mSME, the Yukawa sector is {\it CPT}-even and is given by:
\begin{eqnarray}
\label{YL}
{\cal L}_{Y}&=&- (Y_L)^{AB}\bar{L}_A\phi R_B-\frac{1}{2}(H_L)^{AB}_{\mu \nu}\bar{L}_A\phi \sigma^{\mu \nu} R_B \nonumber \\
&&- (Y_U)^{AB}\bar{Q}_A\tilde{\phi} U_B-\frac{1}{2}(H_U)^{AB}_{\mu \nu}\bar{Q}_A\tilde{\phi} \sigma^{\mu \nu} U_B \nonumber \\
&&- (Y_D)^{AB}\bar{Q}_A\phi D_B-\frac{1}{2}(H_D)^{AB}_{\mu \nu}\bar{Q}_A\phi \sigma^{\mu \nu} D_B +H.c. \, ,
\end{eqnarray}
where $\phi$ is the Higgs doublet; $L(R)$, $Q(U,D)$ are the usual left-handed lepton doublet (right-handed lepton singlet) and the left-handed quark doublet (right-handed quark singlets) of $SU_L(2)$, respectively. The matrices $(H_{(L,U,D)})_{\mu \nu}$ are dimensionless, but as it happens with the SM matrices $Y_{(L,U,D)}$, they are not necessarily Hermitian in flavor space. This opens an window to look for flavor-violation effects mediated by the Higgs boson~\cite{LFVSME}.

Once implemented the standard unitary transformations to pass from the gauge basis to mass-eigenstates basis, the Lagrangian given in Eq.~(\ref{YL}) can be written, in the unitary gauge, as follows:
\begin{equation}
  {\cal L}_Y=-\sum_{A}\left(m_{f_A}+\frac{g\,m_{f_A}}{2m_W}\,H\right)\bar{f}_{A}f_{A}-\frac{1}{2}\sum_{A,B}\left(v+H\right)\bar{f}_A
  \left(V^{AB}_{\alpha \beta}+A^{BA*}_{\alpha \beta}\, \gamma^5\right)\sigma^{\alpha \beta}\,f_B \, ,
\end{equation}
where
\begin{eqnarray}
V^{AB}_{\alpha \beta}&=&\frac{1}{2}\left(Y^{AB}_{\alpha \beta}+Y^{BA*}_{\alpha \beta}\right)\, ,\\
A^{AB}_{\alpha \beta}&=&\frac{1}{2}\left(Y^{AB}_{\alpha \beta}-Y^{BA*}_{\alpha \beta}\right)\, .
\label{AABab}
\end{eqnarray}
In the above expressions,  $Y_{\alpha \beta}=V^\dag_L H_{\alpha \beta} V_R$, with $V_{L,R}$ the standard unitary matrices that connect the gauge and mass eigenstates. Though not explicitly indicated in the matrices $Y_{\alpha \beta}$, it is understood that there are three types of flavor matrices, namely, $Y^{L}_{\alpha \beta}$, $Y^{D}_{\alpha \beta}$, and $Y^{U}_{\alpha \beta}$. In a perturbative approach, which is adopted here, there are two types of physical couplings: the insertion $-(v/2)\bar{f_A}\left(V^{AB}_{\alpha \beta}+A^{BA*}_{\alpha \beta}\, \gamma^5\right)\sigma^{\alpha \beta}f_B$ and the vertex $-(1/2)H\bar{f_A}\left(V^{AB}_{\alpha \beta}+A^{BA*}_{\alpha \beta}\, \gamma^5\right)\sigma^{\alpha \beta}f_B$. At the one-loop level, the contribution to the photon propagator is given by the bilinear term, whose Feynman rule is $-i(v/2)\left(V^{AB}_{\alpha \beta}+A^{BA*}_{\alpha \beta}\, \gamma^5\right)\sigma^{\alpha \beta}$. Note, from Eq.~(\ref{AABab}), that $A_{\alpha \beta}$ vanishes for Hermitian matrices, that is, if $Y^\dag_{\alpha \beta}=Y_{\alpha \beta}$.

\subsection{The Higgs-sector extension}
\label{EHK}
~
As regards the Higgs kinetic term, in the mSME it is extended by {\it CPT}-even\footnote{A {\it CPT}-odd term,  which does not contribute to the photon propagator, is proposed in~\cite{SME}. However, as it is argued in Ref.~\cite{LVG2}, this term may be unobservable since it can be absorbed by a field redefinition.} parts as
\begin{equation}
\label{HKSME}
{\cal L}^{CPT\text{-even}}_{H}=\left[g^{\mu \nu} +(k_{\phi \phi})^{\mu \nu}\right](D_\mu \phi)^\dag (D_\nu \phi)-\frac{1}{2}k^{\mu \nu}_{\phi W}\left(\phi^\dag \frac{\sigma^i}{2} \phi\right)gW^i_{\mu \nu}-\frac{1}{2}k^{\mu \nu}_{\phi B}\left(\phi^\dag\frac{Y}{2}\phi\right) g'B_{\mu \nu}\, .
\end{equation}
In this equation, we have shown the group generators of the electroweak group explicitly. Furthermore, in order for the expressions that will be subsequently derived to be as transparent as possible, we have multiplied each curvature by the coupling constant of the respective group. In the above expressions, the tensors $k_{\phi \phi}^{\mu \nu}$, $k^{\mu \nu}_{\phi W}$, and $k^{\mu \nu}_{\phi B}$ are all dimensionless, being the later two real and antisymmetric, whereas the first one could, in principle, have a real symmetric part and an imaginary antisymmetric part:
\begin{equation}
(k_{\phi \phi})^{\mu \nu}=(k^S_{\phi \phi})^{\mu \nu}+i(k^A_{\phi \phi})^{\mu \nu}\, .
\end{equation}
However, as it was pointed out in~\cite{LVG2} in the context of scalar electrodynamics, there is a redundancy in the count of antisymmetric tensors. In fact, a more careful analysis shows us that only can appear two independent antisymmetric tensors, namely, those associated with the curvatures of the electroweak Yang-Mills sector, just as the last two terms appearing in the (\ref{HKSME}) Lagrangian. To see that the antisymmetric part of $k_{\phi \phi}^{\mu \nu}$ can be moved to the last two terms of (\ref{HKSME}), we use integration by parts to obtain the following relation:
\begin{equation}
i(k^A_{\phi \phi})^{\mu \nu}(D_\mu \phi)^\dag (D_\nu \phi)=-\frac{1}{2}(k^A_{\phi \phi})^{\mu \nu}\left[\left(\phi^\dag \frac{\sigma^i}{2}\phi\right)g W^i_{\mu \nu}+\left(\phi^\dag \frac{Y}{2}\phi\right)g'B_{\mu \nu}\right]\, ,
\end{equation}
which shows us that $(k^A_{\phi \phi})^{\mu \nu}$ is not independent. From now on, we will assume that $k^{\mu \nu}_{\phi \phi}$ is symmetric and real. Using this result, the Lagrangian (\ref{HKSME}) can be written in the following form:
\begin{eqnarray}
{\cal L}^{CPT\text{-even}}_{H}&=&\left(g^{\mu \nu} +k_{\phi \phi}^{\mu \nu}\right)\Big\{\frac{1}{4}\left[\left(\hat{D}_\mu +\bar{D}_\mu \right)G^+_W\right]^\dag \left[\left(\hat{D}_\nu +\bar{D}_\nu \right)G^+_W\right]\nonumber \\
&&+\frac{g^2}{2}\left(\varphi^{0*}\varphi^0 W^-_\mu W^+_\nu +
G^-_WG^+_W W^-_\nu W^+_\mu \right)+ \hat{{\cal L}}_{H}\Big\}+{\cal L}_{\phi WB}\, ,
\end{eqnarray}
where
\begin{eqnarray}
\label{HL}
 \hat{{\cal L}}_{H}&=&\frac{1}{4}\left[\left(\hat{D}^\dag_\nu +\bar{D}_\nu \right)\varphi^0\right]^\dag \left[\left(\hat{D}^\dag_\mu +\bar{D}_\mu \right)\varphi^0\right]+\frac{ig}{2\sqrt{2}}\Big[\varphi^{0*}W^-_\mu\left(\hat{D}_\nu +\bar{D}_\nu\right)G^+_W-\varphi^{0}W^+_\nu\left(\hat{D}_\mu +\bar{D}_\mu\right)^\dag G^-_W \nonumber \\
 &&+ G^-_WW^+_\mu\left(\hat{D}^\dag_\nu+\bar{D}_\nu\right)\varphi^0-G^+_WW^-_\nu\left(\hat{D}_\mu+\bar{D}^\dag_\mu\right)\varphi^{0*} \Big]\, ,
\end{eqnarray}

\begin{eqnarray}
\label{phiWB}
{\cal L}_{\phi WB}&=&-\frac{g}{4}k^{\mu \nu}_{\phi W}\left[\sqrt{2}\varphi^0G^-_W\hat{W}^+_{\mu \nu}+\sqrt{2}\varphi^{0*}G^+_W\hat{W}^-_{\mu \nu} + \left(G^-_WG^+_W-\varphi^{0*}\varphi^0 \right)\hat{W}^3_{\mu \nu}\right]\nonumber \\
&&-\frac{g'}{4}k^{\mu \nu}_{\phi B}\left(G^-_WG^+_W+\varphi^{0*}\varphi^0 \right)B_{\mu \nu}\, .
\end{eqnarray}
In the above expressions, we have introduced the definitions $\varphi^0=(v+H+iG_Z)/\sqrt{2}$, $\hat{D}_\mu=D_\mu-igc_WZ_\mu$, and $\bar{D}_\mu=D_\mu+ig(s^2_W/c_W)Z_\mu$, with $D_\mu=\partial_\mu-ieA_\mu$ the electromagnetic covariant derivative. In addition,  $s_W$ ($c_W$) is the sine (cosine) of the weak angle $\theta_W$, $G^+_W(G_Z)$ is the  pseudo-Goldstone boson associated with the $W^+_\mu(Z_\mu)$ gauge boson, and
\begin{eqnarray}
\hat{W}^+_{\mu \nu}&=&\hat{D}_\mu W^+_\nu-\hat{D}_\nu W^+_\mu \, ,\\
\hat{W}^3_{\mu \nu}&=&s_WF_{\mu \nu}+c_WZ_{\mu \nu}+ig\left(W^-_\mu W^+_\nu- W^+_\mu W^-_\nu\right)\, ,\\
B_{\mu \nu}&=&c_W F_{\mu \nu}-s_W Z_{\mu \nu}\, ,
\end{eqnarray}
where $Z_{\mu \nu}=\partial_\mu Z_\nu-\partial_\nu Z_\mu$.

The physical interactions from the Higgs sector extension can be easily identified in the unitary gauge. In this gauge, we have
\begin{eqnarray}
\label{HUG}
{\cal L}^{\rm UG}_{H}&=&\left(g^{\mu \nu} +k_{\phi \phi}^{\mu \nu}\right)\left[\frac{1}{2}(\partial_\mu H)(\partial^\mu H)+\frac{1}{2}m^2_Z\left(1+\frac{H}{v}\right)^2Z_\mu Z_\nu +m^2_W\left(1+\frac{H}{v}\right)^2W^{-}_\mu W^{+}_\nu\right]\nonumber \\
&&+\frac{g}{8}(v+H)^2\left(k_{\phi W}^{\mu \nu}\hat{W}^3_{\mu \nu}-\frac{s_W}{c_W}k_{\phi B}^{\mu \nu}B_{\mu \nu}\right)\, .
\end{eqnarray}
We can see that, in this gauge, the only contribution from $(k_{\phi \phi})_{\mu \nu}$ to the photon propagator arises from the modification to the $W$ propagator given by $k_{\phi \phi}^{\mu \nu}m^2_W\,W^{-}_\mu W^{+}_\nu$, whereas $k^{\mu \nu}_{\phi W}$ contributes, in addition, through a vertex $A_\mu H H$. In this gauge, the coefficient  $k^{\mu \nu}_{\phi B}$ can contribute to the photon propagator only through the vertex $A_\mu H H$, while additional contributions emerge via pseudo-Goldstone bosons if an $R_\xi$-gauge is chosen. Note that the antisymmetric tensors $k^{\mu \nu}_{\phi W}$, and $k^{\mu \nu}_{\phi B}$ generate a finite tree-level contribution to the photon propagator through a $A_\mu H$ mixing, but such a contribution can be absorbed by a redefinition of the electromagnetic field.

\subsection{The gauge-sector extension}
The gauge-sector extension includes both $CPT$-even and $CPT$-odd effects. Here, we focus only on the $CPT$-even part, which is given by:
\begin{equation}
{\cal L}^{\rm EW}_\textrm{gs}=-\frac{1}{4}\left[g_{\lambda \mu}g_{\rho \nu}+(k_W)_{\lambda \rho \mu \nu}\right]W^{i\lambda \rho}W^{i\mu \nu}-\frac{1}{4}\left[g_{\lambda \mu}g_{\rho \nu}+(k_B)_{\lambda \rho \mu \nu}\right]B^{\lambda \rho}B^{\mu \nu}\, ,
\end{equation}
where the SM part has been included. Passing to the mass-eigenstates basis, one has
\begin{eqnarray}
\label{gs}
{\cal L}^{\rm EW}_\textrm{gs}&=&-\frac{1}{4}F_{\mu \nu}F^{\mu \nu}-\frac{1}{4}(k_F)_{\lambda \rho \mu \nu}F^{\lambda \rho}F^{\mu \nu}
-\frac{1}{4}Z_{\mu \nu}Z^{\mu \nu}-\frac{1}{4}(k_Z)_{\lambda \rho \mu \nu}Z^{\lambda \rho}Z^{\mu \nu}\nonumber \\
&&-\frac{1}{4}\frac{s_{2W}}{c_{2W}}(k_F-k_Z)_{\lambda \rho \mu \nu}F^{\lambda \rho}Z^{\mu \nu}-\frac{1}{2}\left[g_{\lambda \mu}g_{\rho \nu}+(k_W)_{\lambda \rho \mu \nu}\right]\Big\{\hat{W}^{-\lambda \rho}\hat{W}^{+\mu \nu}\nonumber \\
&&+ig\left[2\left(s_WF^{\lambda \rho}+c_WZ^{\lambda \rho}\right)+ig\left(W^{-\lambda}W^{+\rho}-W^{+\lambda}W^{-\rho}\right)\right]W^{-\mu}W^{+\nu}
\Big\}\, ,
\end{eqnarray}
where $k_F=s^2_Wk_W+c^2_Wk_B$ and $k_Z=c^2_Wk_W+s^2_Wk_B$. In addition, $s_{2W} (c_{2W})$ stands for the sine (cosine) of $2\theta_W$. Notice that we have three Riemann-type tensors, namely, $k_W$, $k_Z$, and $k_F$. Observe that the $A-Z$ mixing appearing in (\ref{gs}) leads to a tree-level contribution to the photon propagator proportional to $(k_F-k_Z)_{\alpha \beta \gamma \delta}(k_F-k_Z)_{\lambda \rho \mu \nu}g^{\delta \nu}\partial^\gamma F^{\alpha \beta}\partial^\mu F^{\lambda \rho}$, which, however, is very suppressed by birefringence.

The photon propagator is, as most off-shell Green's functions, gauge dependent, and the unitary gauge is not the most appropriate choice to carry out this calculation. It is, rather, convenient to use a $U_Q(1)$-covariant nonlinear gauge~\cite{NLG}, which considerably simplifies the calculations. In the SM part, this gauge removes, among others, the unphysical vertex $WG_W\gamma$ and the bilinear mixing $WG_W$. These unphysical interactions are also induced in the mSME, but they remain in the theory even if this gauge is utilized. Due to this, it is desirable, from a practical viewpoint, to introduce some sort of generalization of this type of gauge that includes LV without affecting the predictive structure of the theory. This will naturally introduce LV in the ghost sector. Even though the ghost sector emerges as a consequence of introducing a gauge-fixing procedure, its presence is essential because the ghost fields are needed to quantize the theory. In the case of Abelian theories, the issue of LV in the ghost sector has already been addressed in~\cite{LVG1,LVG2} within the context of the Faddeev-Popov method (FPM)~\cite{FP}. In the next subsection, we will address this problem in the context of the mSME from the perspective of the BRST symmetry. As we commented in the Introduction, our main purpose is to introduce a gauge-fixing procedure that simplifies as much as possible the calculations of radiative corrections in the mSME.

\section{A Nonlinear gauge for the SME}
\label{GFP}
The structure of the ghost sector is dictated by the gauge-fixing procedure used to quantize the theory. The cornerstone of any gauge-fixing procedure are the so-called gauge-fixing functions, which depend, in general, on gauge and scalar fields, and are, by construction, Lorentz invariant but not covariant objets in the sense of gauge transformations\footnote{Nevertheless, a sort of gauge invariant gauge-fixing procedure can be introduced for theories with and without spontaneous symmetry breaking through the Background Field Method~\cite{BFM}.}. Their fundamental role is the definition of the gauge propagators, which cannot be carried out as long as gauge invariance holds. In theories with spontaneous symmetry breaking, these functions are judiciously chosen to define unphysical masses for the pseudo-Goldstone bosons and ghost fields, as well as to remove bilinear couplings between gauge fields and pseudo-Goldstone bosons from the Higgs sector. Such gauges allow us to define well-behaved propagators for massive gauge fields in the ultraviolet range. Linear gauge-fixing functions in both types of fields are adequate to this purpose~\cite{LG}. However, renormalization theory allows usage of gauge-fixing functions that are quadratic with respect to gauge and/or scalar fields, thus defining {\it nonlinear gauges}. The introduction of nonlinear-gauge functions can modify the vertex structure of the gauge sector of the theory in a nontrivial way, with changes in the ghost sector taking place as well. Because in theories with spontaneous symmetry breaking the Higgs sector is an essential piece of the gauge structure of the theory, this sector can be profoundly altered when a nonlinear gauge is introduced~\cite{NLG}. Indeed, the main motivation to introduce nonlinear gauges is to simplify this sector as much as possible, since it involves unphysical vertices that survive linear gauges. Our main purpose in this section is to define a nonlinear gauge-fixing procedure aimed at simplifying the Higgs sector of the mSME as much as possible.

\subsection{BRST symmetry and gauge-fixing procedures}
It is a known fact that the FPM fails to quantize Yang-Mills theories using general linear gauge-fixing functions. The main problem has to do with renormalization. The reason behind this is that the FPM leads to an action which is bilinear in the ghost and antighost fields as they emerge essentially from the integral representation of a determinant. This is not the most general situation, because an action including quartic ghost interactions is still consistent with the symmetries of the theory and the power counting criterion of renormalization theory. It turns out that, within the context of the FPM, quartic ghost interactions cannot arise from radiative corrections because the theory is invariant under translations $\bar{C}^a \to \bar{C}^a+c^a$, with $\bar{C}^a$ the antighost field and $c^a$ a constant anticommuting number. This symmetry arises as a consequence of the fact that the antighost fields only appear through their derivatives, which, however, is no longer true in more general gauge-fixing procedures, such as nonlinear ones since in this case the gauge-fixing functions depend quadratically on the gauge fields. The presence of these terms lead to divergent quartic ghost interactions at one loop, which means that renormalizability gets spoiled when the FPM is attempted in the context of nonlinear gauges. Instead of using this method to construct the ghost sector of the mSME, we will resort to BRST symmetry~\cite{BRST1,BRST2,BRST3}, which has proved to be a powerful tool that is adequate not only to quantize Yang-Mills theories with broader gauge-fixing procedures, as nonlinear ones, but also to quantize more-general gauge systems~\cite{BRST3}. Due to the central role played by this symmetry in our calculations, we provide, in the context of the SM, a brief review of its main properties in Appendix~\ref{FAFF}.\\

Our starting point is the gauge-fixed BRST action appearing in the path integral,
\begin{equation}
\label{EQA}
S_{\rm eff}=\int d^4x\left\{{\cal L}_{\rm SME}+{\cal L}_{\rm GF}+{\cal L}_{\cal C}  \right\}\, ,
\end{equation}
where ${\cal L}_{\rm GF}$ is the gauge-fixing Lagrangian, which is given by
\begin{equation}
{\cal L}_{\rm GF}=-\frac{1}{2\hat{\xi}}f^af^a-\frac{1}{2\xi}f^if^i-\frac{1}{2\xi}f^2\, ,
\end{equation}
whereas the Lagrangian for the ghost sector, ${\cal L}_{\cal C}$, can be written as:
\begin{equation}
{\cal L}_{\cal C}={\cal L}_{{\cal C}\rm G}+{\cal L}_{{\cal C} {\cal C}}+{\cal L}_{\rm GH}\, ,
\end{equation}
with
\begin{equation}
{\cal L}_{{\cal C}\rm G}=\bar{\cal C}^a \frac{\partial f^a}{\partial G^b_\mu}{\cal D}^{bc}_\mu {\cal C}^c+\bar{\cal C}^i \frac{\partial f^i}{\partial W^j_\mu}{\cal D}^{jk}_\mu {\cal C}^k+\bar{\cal C}\frac{\partial f}{\partial B_\mu}\partial_\mu {\cal C}\, ,
\end{equation}
\begin{eqnarray}
{\cal L}_{{\cal C} {\cal C}}&=&-\frac{1}{\hat{\xi}}f^{abc}f^a\bar{\cal C}^b {\cal C}^c+\frac{1}{2}f^{abc}f^{ade}\left(\frac{1}{\hat{\xi}}\, \bar{\cal C}^b\bar{\cal C}^d{\cal C}^c{\cal C}^e+ \bar{\cal C}^d\bar{\cal C}^e{\cal C}^b{\cal C}^c \right)\nonumber \\
&&-\frac{1}{\xi}\epsilon^{ijk}f^i\bar{\cal C}^j {\cal C}^k+\frac{1}{2}\epsilon^{ijk}\epsilon^{ilm}\left(\frac{1}{\xi}\, \bar{\cal C}^j\bar{\cal C}^l{\cal C}^k{\cal C}^m+ \bar{\cal C}^l\bar{\cal C}^m{\cal C}^j{\cal C}^k \right)\, ,
\end{eqnarray}
while the ${\cal L}_{\rm GH}$ Lagrangian contains the ghost-Higgs interactions, which is given in~(\ref{GHi}). In the above expressions, $f^a$, $f^i$, and $f$ are the gauge-fixing functions associated with the $SU_C(3)$, $SU_L(2)$, and $U_Y(1)$ gauge groups, respectively (see Appendix~\ref{FAFF} for details).

We now introduce a gauge-fixing procedure by defining the functions $f^a$, $f^i$, and $f$. From our discussion in Appendix~\ref{FAFF} on the main properties of the gauge-fixing fermion $\Psi$, it has become clear that this set of gauge-fixing functions provide us with the only mechanism that allows us to introduce LV in the ghost sector without violating the BRST symmetry. Also, from the renormalizable structure of the functional $\Psi$, we note that the gauge-fixing functions must be real and satisfy the power-counting criterion of renormalization theory, which in turn implies that they must depend on gauge and scalar fields, at most, quadratically. Although, in general, no symmetry criteria to construct specific gauge-fixing functions exist\footnote{However, see~\cite{BFM} and references therein.}, there is an exception in the case of theories with spontaneous symmetry breaking. When a theory characterized by a gauge group $G$ is broken down into one of its nontrivial subgroups $H$ at certain energy scale, some of the gauge fields associated with the broken generators of $G$ are mapped into a tensorial representation of $H$. In such a situation, the introduction of a gauge-fixing procedure for these gauge fields that is covariant under the $H$ group turns out to be convenient~\cite{NLGG}. This class of gauges are nonlinear because they involve covariant derivatives instead of simple derivatives $\partial_\mu$. In the SM, the electroweak group $SU_L(2)\times U_Y(1)$ is broken at the Fermi scale into the electromagnetic group $U_Q(1)$, and this scenario is not altered when LV is introduced. In this case, it is convenient to implement a $U_Q(1)$-covariant gauge-fixing procedure for the gauge boson $W^\pm$. With this in mind, we generalize the nonlinear gauge for the SM, given in~\cite{NLG}, to include LV. Because the mSME incorporates new physical parameters characterized by constant Lorentz tensors, we introduce gauge-fixing functions that include, in addition, some of these parameters. However, keep in mind that, as in the SM case, these extended gauge-fixing functions only can depend on physical parameters that are already present in the classical action; no new parameters can be introduced. How many and what kind of parameters should we introduce is a matter of convenience that, in a practical sense, depends on our ability to simplify the Higgs sector and the gauge sector of the mSME as much as possible. Moreover, strictly speaking, nonlinear gauges are not necessary to quantize the theory, neither in the SM nor in the SME, so we emphasize that the introduction of such gauges is not a fundamental necessity, but rather a practical subject that aims at rendering calculations simpler. Usual linear gauges, which are contained in the nonlinear ones, are indeed sufficient for this purpose. From now on, we focus only on the electroweak sector.

 In the case of the $SU_L(2)$ group, we introduce a generalization of the nonlinear gauge given in~\cite{NLG}. In this case, the gauge-fixing functions $f^i$ depend on both gauge and scalar fields, so it is convenient to decompose them into a vector part plus a scalar part, as follows:
\begin{equation}
f^i=f^i_\textrm{v}+f^i_\textrm{s}\, .
\end{equation}
A choice for the vector term that includes LV, which considerably simplifies the Higgs and gauge sectors, is
\begin{equation}
\label{V1}
f^i_\textrm{v}=\left(g^{\mu \nu}+k_{\phi \phi}^{\mu \nu}\right)\left(\mathfrak{D}^{ij}_\mu W^j_\nu \right)\, ,
\end{equation}
where $\mathfrak{D}^{ij}_\mu=\delta^{ij}\partial_\mu-g'\epsilon^{3ij}B_\mu$. Note that this object contains the electromagnetic covariant derivative. It is important to stress that it is not possible to introduce either of the two antisymmetric tensors $k_{\phi W}^{\mu \nu}$ or $k_{\phi B}^{\mu \nu}$ in Eq.(\ref{V1}) because to perform any useful cancelation in the Lagrangian (\ref{phiWB}) it would be necessary to introduce these tensor together a factor $i$, which is in conflict with the fact that the gauge-fixing functions are restricted to be real (see Appendix~\ref{FAFF}).

As far as the scalar part $f^i_\textrm{s}$ is concerned, a judicious nonlinear choice would allow us to eliminate a significant number of unphysical vertices that appear in the Higgs sector, in addition to defining unphysical masses for the pseudo-Goldstone bosons and ghosts. With this in mind, we introduce the following scalar gauge-fixing functions:
\begin{equation}
f^i_\textrm{s}=\frac{ig\xi}{2}\left[\phi^\dag\left(\sigma^i-i\epsilon^{3ij}\sigma^j\right)\phi_0-
\phi_0^\dag\left(\sigma^i+i\epsilon^{3ij}\sigma^j\right)\phi+i\epsilon^{3ij}\phi^\dag \sigma^j\phi \right]\, ,
\end{equation}
where $\phi^\dag_0=(0,\frac{v}{\sqrt{2}})$. Besides effects generated through its interference with the vector part, this type of scalar gauge-fixing functions modify the Higgs potential. Note that no LV is considered in this part, since in the mSME the Higgs potential does not include it.

With regard to the  hypercharge group $U_Y(1)$, we also write the corresponding gauge-fixing function as the sum of a vector part and a scalar part:
\begin{equation}
f=f_\textrm{v}+f_\textrm{s}\, ,
\end{equation}
where
\begin{eqnarray}
f_\textrm{v}&=&\left(g^{\mu \nu}+k_{\phi \phi}^{\mu \nu}\right)(\partial_\mu B_\nu) \, ,\\
f_\textrm{s}&=&\frac{ig'\xi}{2} \left(\phi^\dag \phi_0- \phi^\dag_0 \phi\right)\, ,
\end{eqnarray}
Putting the $k_{\phi \phi}^{\mu \nu}$ tensor equal to zero, we recover the standard nonlinear gauge given in~\cite{NLG}. If, in addition, we introduce the changes $\mathfrak{D}^{ij}_\mu \to \delta^{ij}\partial_\mu$ and $\epsilon^{3ij}\to 0$, we recover the well-known linear gauge~\cite{LG}.

\subsection{Implications on the gauge and Higgs sectors }
 The main feature of the above gauge-fixing procedure is that the $W^\pm$-gauge-boson propagator is defined in a covariant way under the electromagnetic gauge group. We now proceed to study the corresponding implications on the bosonic sector of the mSME. Using the standard relations $f^\pm=\frac{1}{\sqrt{2}}(f^1\mp if^2)$, $f^Z=c_Wf^3-s_Wf$, and $f^A=s_Wf^3+c_Wf$ to pass from the gauge basis, characterized by gauge-fixing functions $f^1,f^2,f^3,f$, to the mass basis, in which the gauge fixing-functions are denoted by $f^-,f^+,f^Z,f^A$, we write the gauge-fixing sector as
 \begin{eqnarray}
 {\cal L}^{\rm GF}_\textrm{vv}&=&-\frac{1}{\xi}f^-_\textrm{v}f^+_\textrm{v}-\frac{1}{2\xi}f^Z_\textrm{v}f^Z_\textrm{v}
 -\frac{1}{2\xi}f^A_\textrm{v}f^A_\textrm{v}\, ,\\
 {\cal L}^{\rm GF}_\textrm{vs}&=&-\frac{1}{\xi}\left(f^-_\textrm{v}f^+_\textrm{s}+f^+_\textrm{v}f^-_\textrm{s} \right)-\frac{1}{\xi}f^Z_\textrm{v}f^Z_\textrm{s}\, ,  \\
 {\cal L}^{\rm GF}_\textrm{ss}&=&-\frac{1}{\xi}f^-_\textrm{s}f^+_\textrm{s}-\frac{1}{2\xi}f^Z_\textrm{s}f^Z_\textrm{s}\, ,
 \end{eqnarray}
where
\begin{equation}
\label{gfvv}
 {\cal L}^{\rm GF}_\textrm{vv}=-\frac{1}{\xi}\left(g^{\mu \nu}+k_{\phi \phi}^{\mu \nu}\right)\left(g^{\lambda \rho}+k_{\phi \phi}^{\lambda \rho}\right)
\left[(\bar{D}_\mu W^+_\nu)^\dag(\bar{D}_\lambda W^+_\rho)+\frac{1}{2}\left(\partial_\mu Z_\nu\partial_\lambda Z_\rho+
\partial_\mu A_\nu\partial_\lambda A_\rho \right) \right]\, ,
\end{equation}
\begin{equation}
{\cal L}^{\rm GF}_\textrm{vs}=\left(g^{\mu \nu}+k_{\phi \phi}^{\mu \nu}\right)\left\{ig\left[\varphi^{0*}G^+_W(\bar{D}_\mu W^+_\nu)^\dag
-\varphi^{0}G^-_W(\bar{D}_\mu W^+_\nu)\right]+m_ZG_Z(\partial_\mu Z_\nu)\right\}\, ,
\end{equation}
\begin{equation}
 {\cal L}^{\rm GF}_\textrm{ss}=-\xi m^2_W\left[\left(1+\frac{g}{2m_W}H\right)^2+\frac{g^2}{4m^2_W}G^2_Z\right]G^-_WG^+_W-\frac{1}{2}\xi m^2_Z G^2_Z\, .
\end{equation}
The above Lagrangians introduce nontrivial modifications in the bosonic part of the electroweak sector of the SME. The modified gauge sector is obtained by adding Eqs.~(\ref{gs}) and (\ref{gfvv}) together,
\begin{equation}
\label{VVYM}
{\cal L}^{\rm EW}_\textrm{gauge}={\cal L}^{\rm EW}_\textrm{gs}+{\cal L}^{\rm GF}_\textrm{vv}\, .
\end{equation}
 On the other hand, the ${\cal L}^{\rm GF}_\textrm{vs}$ Lagrangian introduces important simplifications in the symmetric part of the Higgs sector characterized by the Lagrangian $\hat{\cal L}_H$ (see Eq.(\ref{HL})). In fact, after an integration by parts to remove bilinear terms, we obtain
\begin{eqnarray}
{\cal L}^{\rm GF}_\textrm{vs}+\left( g^{\mu \nu}+k_{\phi \phi}^{\mu \nu}\right)\hat{\cal L}_H&=&\frac{g}{2}\left(g^{\mu \nu}+k_{\phi \phi}^{\mu \nu}\right)\Big\{i\left(G^-_WW^+_\nu \partial_\mu \varphi^0-G^+_WW^-_\nu \partial_\mu \varphi^{0*} \right) \nonumber \\
&&+\partial_\mu \varphi^{0*} \partial_\nu \varphi^0+\frac{g^2}{4c^2_W}\varphi^{0*}\varphi^0 Z_\mu Z_\nu
+\frac{g}{c_W}Z_\mu \left(H\partial_\nu G_Z-G_Z\partial_\nu H  \right) \Big\}\, .
\end{eqnarray}
We can see that the unphysical vertices $WG_W$, $ZG_Z$, $WG_W\gamma$, $WG_WZ$, $HWG_W\gamma$, $HWG_WZ$, $G_ZWG_W\gamma$, and $G_ZWG_WZ$, which are part of the sole Lagrangian term $\hat{\cal L}_H$, have been removed from the Higgs sector. Note that these vertices are canceled in both the SM part and the Lorentz-violating part proportional to $k_{\phi \phi}^{\mu \nu}$. The advantages of using a $U_Q(1)$-covariant gauge can now be appreciated. On the other hand, ${\cal L}^{\rm GF}_\textrm{ss}$ modifies the Higgs potential, but we do not present the corresponding expressions here.

We now focus on the electromagnetic couplings of the $W$ gauge boson, which we need to carry out our calculations. The $W$ quadratic terms that arise from the Lagrangians given by Eqs. (\ref{HUG}) and (\ref{VVYM}) are:
\begin{eqnarray}
\label{TP}
{\cal L}_{WW}&=&-\frac{1}{2}W^-_{\mu \nu}W^{+\mu \nu}-\frac{1}{\xi}(\partial_\mu W^{-\mu})(\partial_\nu W^{+\nu})+m^2_WW^-_\mu W^{+\mu}-\frac{1}{2}(k_W)_{\lambda \rho \mu \nu}W^{-\lambda \rho}W^{+\mu \nu}+im^2_W(k_{\phi W})_{\mu \nu}W^{-\mu }W^{+\nu}\nonumber \\
&&-\frac{1}{\xi}\left[g_{\mu \nu}(k_{\phi \phi})_{\lambda \rho}+g_{\lambda \rho}(k_{\phi \phi})_{\mu \nu}+(k_{\phi \phi})_{\mu \nu}(k_{\phi \phi})_{\lambda \rho} \right](\partial^\mu W^{-\nu})(\partial^\lambda W^{+\rho})+(k_{\phi \phi})_{\mu \nu}m^2_WW^{-\mu}W^{+\nu}\, ,
\end{eqnarray}
where $W^+_{\mu \nu}=\partial_\mu W^+_\nu-\partial_\nu W^+_\mu$. The first three terms of (\ref{TP}) determine the standard $W$ propagator of renormalizable type, whereas the remaining terms can be considered, from the perspective of Feynman diagrams, as insertions. Below we will need the two-point vertex function associated with the ${\cal L}_{WW}$ Lagrangian, which we will denote by $i\Gamma^{WW}_{\mu \nu}(k)$ (see Table~\ref{FR} for notation and conventions), where
\begin{equation}
\Gamma^{WW}_{\mu \nu}(k)=\left(k^2+m^2_W\right)g_{\mu \nu}-\left(1-\frac{1}{\xi}\right)k_\mu k_\nu+\hat{\Gamma}^{WW}_{\mu \nu}(k)+\bar{\Gamma}^{WW}_{\mu \nu}(k)\, ,
\end{equation}
with
\begin{eqnarray}
\label{VFWW1}
\hat{\Gamma}^{WW}_{\mu \nu}(k)&=&-m^2_W\left(k_{\phi \phi}+ik_{\phi W} \right)_{\mu \nu}+2(k_W)_{\lambda \mu \rho \nu}k^\lambda k^\rho \, , \\
\label{VFWW2}
\bar{\Gamma}^{WW}_{\mu \nu}(k)&=&\frac{1}{\xi}\left[g_{\nu \rho}(k_{\phi \phi})_{\lambda \mu}+g_{\mu \rho}(k_{\phi \phi})_{\lambda \nu}+ (k_{\phi \phi})_{\lambda \mu}(k_{\phi \phi})_{\rho \nu}\right]k^\lambda k^\rho \, ,
\end{eqnarray}
On the other hand, the cubic, $W^-W^+\gamma$, and quartic, $W^-W^+\gamma \gamma$, vertices are given by:
\begin{eqnarray}
{\cal L}_{WW\gamma}&=&ie\Big\{\left[g_{\lambda \mu}g_{\rho \nu}+(k_W)_{\lambda \rho \mu \nu} \right]\left[\left(W^{+\lambda \rho}W^{-\mu}- W^{-\lambda \rho}W^{+\mu}\right)A^\nu-F^{\lambda \rho}W^{-\mu}W^{+\nu} \right]\nonumber \\
&&-\frac{1}{\xi}\left[g_{\mu \nu}+(k_{\phi \phi})_{\mu \nu}\right]\left[g_{\lambda \rho}+(k_{\phi \phi})_{\lambda \rho}\right]\left(
A^\mu W^{-\nu}\partial^\lambda W^{+\rho}-A^\lambda W^{+\rho}\partial^\mu W^{-\nu} \right)
  \Big\}\, ,
\end{eqnarray}
\begin{eqnarray}
{\cal L}_{WW\gamma \gamma}&=&-e^2\Big\{\frac{1}{2}\left[g_{\lambda \mu}g_{\rho \nu}+(k_W)_{\lambda \rho \mu \nu} \right]\left(W^{-\lambda}A^\rho-W^{-\rho}A^\lambda \right)\left(W^{+\mu}A^\nu-W^{+\nu}A^\mu \right)\nonumber \\
&&+\frac{1}{\xi}\left[g_{\mu \nu}+(k_{\phi \phi})_{\mu \nu}\right]\left[g_{\lambda \rho}+(k_{\phi \phi})_{\lambda \rho}\right]A^\mu A^\lambda W^{-\nu}W^{+\rho}
\Big\}\, .
\end{eqnarray}
The corresponding vertex functions are denoted by $-ie\Gamma^{WW\gamma}_{\lambda \rho \eta}(k_1,k_2,k_3)$ and $-ie^2\Gamma^{WW\gamma \gamma}_{\alpha \beta\lambda \rho }$, where
\begin{equation}
\Gamma^{WW\gamma}_{\lambda \rho \eta}(k_1,k_2,k_3)=\hat{\Gamma}^{WW\gamma}_{\lambda \rho \eta}(k_1,k_2,k_3)+\bar{\Gamma}^{WW\gamma}_{\lambda \rho \eta}(k_1,k_2,k_3)\, ,
\end{equation}
with
\begin{eqnarray}
\label{WWA1}
\hat{\Gamma}^{WW\gamma}_{\lambda \rho \eta}(k_1,k_2,k_3)&=&\left(k_3-k_2\right)_\eta g_{\lambda \rho}+\left(k_2-k_1+\frac{1}{\xi}k_3\right)_\rho g_{\lambda \eta}+\left(k_1-k_3-\frac{1}{\xi}k_2\right)_\lambda g_{\rho \eta}\nonumber \\
&&+2\left[(k_W)_{\alpha \eta \lambda \rho}k^\alpha_1+ (k_W)_{\alpha \lambda \rho \eta}k^\alpha_2+(k_W)_{\alpha \rho \eta \lambda}k^\alpha_3 \right]\, ,
\end{eqnarray}
\begin{eqnarray}
\label{WWA2}
\bar{\Gamma}^{WW\gamma}_{\lambda \rho \eta}(k_1,k_2,k_3)&=&\frac{1}{\xi}\Big[g_{\lambda \eta}(k_{\phi \phi})_{\rho \alpha}k^\alpha_3+(k_{\phi \phi})_{\lambda \eta}k_{3\rho}+(k_{\phi \phi})_{\lambda \eta} (k_{\phi \phi})_{\rho \alpha}k^\alpha_3\nonumber \\
&&\ \ \ \ -g_{\rho \eta}(k_{\phi \phi})_{\lambda \alpha}k^\alpha_2-(k_{\phi \phi})_{\rho \eta}k_{2\lambda}-(k_{\phi \phi})_{\rho \eta} (k_{\phi \phi})_{\lambda \alpha}k^\alpha_2\Big]\, .
\end{eqnarray}
On the other hand,
\begin{equation}
\Gamma^{WW\gamma \gamma}_{\alpha \beta \lambda \rho}=\hat{\Gamma}^{WW\gamma \gamma}_{\alpha \beta \lambda \rho}+\bar{\Gamma}^{WW\gamma \gamma}_{\alpha \beta \lambda \rho}\, ,
\end{equation}
where
\begin{eqnarray}
\label{WWAA1}
\hat{\Gamma}^{WW\gamma \gamma}_{\alpha \beta \lambda \rho}&=&2g_{\alpha \beta}g_{\lambda \rho}-\left(1-\frac{1}{\xi}\right)\left(g_{\alpha \lambda}g_{\beta \rho}+g_{\alpha \rho}g_{\beta \lambda}\right)+2\left[(k_W)_{\alpha \rho \beta \lambda}+(k_W)_{\beta \rho \alpha \lambda}  \right]\, , \\
\label{WWAA2}
\bar{\Gamma}^{WW\gamma \gamma}_{\alpha \beta \lambda \rho}&=&\frac{1}{\xi}\Big[g_{\alpha \lambda}(k_{\phi \phi})_{\beta \rho}+g_{\beta \lambda}(k_{\phi \phi})_{\alpha \rho} +(k_{\phi \phi})_{\alpha \lambda}(k_{\phi \phi})_{\alpha \rho}+
g_{\alpha \rho}(k_{\phi \phi})_{\beta \lambda}+g_{\beta \rho}(k_{\phi \phi})_{\alpha \lambda} +(k_{\phi \phi})_{\alpha \rho}(k_{\phi \phi})_{\beta \lambda}    \Big]\, .
\end{eqnarray}

It is easy to show that the above vertex functions satisfy the following simple Ward identity:
\begin{equation}
k_1^\eta \Gamma^{WW\gamma}_{\lambda \rho \eta}(k_1,k_2,k_3)=\Gamma^{WW}_{\lambda \rho}(k_2)-\Gamma^{WW}_{\lambda \rho}(k_3)\, ,
\end{equation}
which reflect the $U_Q(1)$-symmetric structure of the gauge-fixing procedure used for the $W$ gauge boson.

In the case of the charged scalar sector, the quadratic, cubic, and quartic couplings are given by
\begin{eqnarray}
{\cal L}_{G_WG_W}&=&\left[g^{\mu \nu}+(k_{\phi \phi})^{\mu \nu}\right](\partial_\mu G^-_W)(\partial_\nu G^+_W)-\xi m^2_W G^-_WG^+_W \, ,\\
{\cal L}_{G_WG_W\gamma}&=&ie\left[g^{\mu \nu}+(k_{\phi \phi})^{\mu \nu}\right]\left(A_\mu G^-_W\partial_\nu G^+_W-A_\nu G^+_W\partial_\mu G^-_W\right)
-\frac{e}{4}\left(k_{\phi W}+k_{\phi B}\right)^{\mu \nu}G^-_WG^+_W F_{\mu \nu}\, ,\\
{\cal L}_{G_WG_W\gamma\gamma}&=&e^2\left[g^{\mu \nu}+(k_{\phi \phi})^{\mu \nu}\right]A_\mu A_\nu G^-_WG^+_W\, ,
\end{eqnarray}
being the respective vertex functions $i\Gamma^{G_WG_W}(k)$, $-ie\Gamma^{G_WG_W\gamma}_\mu (p_1,p_2,q)$, and $ie^2\Gamma^{G_WG_W\gamma\gamma}_{\mu \nu}$, where
\begin{eqnarray}
\Gamma^{G_WG_W}(k)&=&-k^2-\xi m^2_W-(k_{\phi \phi})_{\mu \nu}k^\mu k^\nu \, ,\\
\label{GGA}
\Gamma^{G_WG_W\gamma}_\mu(p_1,p_2,q)&=&\left[g_{\mu \nu}+(k_{\phi \phi})_{\mu \nu}\right](p_1-p_2)^\nu -
\frac{i}{2}\left(k_{\phi W}+k_{\phi B}\right)_{\mu \nu}q^\nu\, , \\
\label{GGAA}
\Gamma^{G_WG_W\gamma\gamma}_{\mu \nu}&=&2\left[g_{\mu \nu}+(k_{\phi \phi})_{\mu \nu}\right]\, .
\end{eqnarray}
It is easy to see that, as it occurs for the case of the $W$ gauge boson, the following simple Ward identity hold:
\begin{equation}
q^\mu \Gamma^{G_WG_W\gamma}_\mu(p_1,p_2,q)=\Gamma^{G_WG_W}(p_1)-\Gamma^{G_WG_W}(p_2)\, ,
\end{equation}

\subsection{The ghost Lagrangian}
The nonlinear gauge introduced above leads to a ghost sector that differs substantially from the one that emerges in the context of the FPM. Using the standard relations for the ghost fields ${\cal C}^\pm=\frac{1}{\sqrt{2}}({\cal C}^1\mp i{\cal C}^2)$, ${\cal C}^Z=c_W{\cal C}^3-s_W{\cal C}$, ${\cal C}^A=s_W{\cal C}^3+c_W{\cal C}$, and similar expressions for the anti-ghost fields, the ghost-gauge Lagrangian for the electroweak sector, ${\cal L}^{\rm EW}_{{\cal C}\rm G}$, can be written as follows:
\begin{eqnarray}
{\cal L}^{\rm EW}_{{\cal C}\rm G}&=&-\left(g^{\alpha \beta}+k_{\phi \phi}^{\alpha \beta}\right)\Big\{ (\bar{D}_\alpha \bar{\cal C}^+)(\hat{D}_\beta {\cal C}^+)^\dag+ (\bar{D}_\alpha \bar{\cal C}^+)^\dag(\hat{D}_\beta {\cal C}^+)+\partial_\alpha \bar{\cal C}^Z\partial_\beta {\cal C}^Z+\partial_\alpha \bar{\cal C}^A\partial_\beta {\cal C}^A\nonumber \\
&&+\left[W^{+}_\beta(\bar{D}_\alpha \bar{\cal C}^+)^\dag- W^{-}_\beta(\bar{D}_\alpha \bar{\cal C}^+) \right]\left(ie\, {\cal C}^A+igc_W{\cal C}^Z\right)+\left(ie\, \partial_\alpha\bar{\cal C}^A+ igc_W\partial_\alpha \bar{\cal C}^Z\right)\left(W^-_\beta {\cal C}^+- W^+_\beta {\cal C}^-\right)
\Big\}\, .
\end{eqnarray}
Notice that this Lagrangian is manifestly invariant under the $U_Q(1)$ gauge group. This means that, in contrast with the case of the FPM, the charged ghost sector obeys Ward identities that resemble those appearing in scalar electrodynamics. This fact leads to important simplifications in radiative corrections.

As far as the ghost-Higgs Lagrangian ${\cal L}_{{\cal C}\rm H}$ is concerned, it is convenient to decompose it into the following two parts:
\begin{equation}
{\cal L}_{{\cal C}\rm H}={\cal L}^l_{{\cal C}\rm H}+{\cal L}^{nl}_{{\cal C}\rm H}\, ,
\end{equation}
where ${\cal L}^l_{{\cal C}\rm H}$ is the well-known result that emerges from a linear gauge, which is given by
\begin{eqnarray}
{\cal L}^l_{{\cal C}\rm H}&=&\frac{g^2\xi}{4}\left[\left(\phi^\dag_0\phi+\phi^\dag\phi_0\right)\delta^{ij}+i\epsilon^{ijk} \left(\phi^\dag_0\sigma^k \phi+\phi^\dag\sigma^k\phi_0\right) \right]\bar{\cal C}^i {\cal C}^j\nonumber \\
&&+\frac{gg'\xi}{4}\left(\phi^\dag_0\sigma^i \phi+\phi^\dag\sigma^i\phi_0\right)\left(\bar{\cal C}^i{\cal C}+ \bar{\cal C}{\cal C}^i\right)+
\frac{g'^2\xi}{4}\left(\phi^\dag_0\phi+\phi^\dag\phi_0\right)\bar{\cal C}{\cal C}\, ,
\end{eqnarray}
whereas ${\cal L}^{nl}_{{\cal C}\rm H}$, which contains the nonlinear effects, can be written as follows:
\begin{eqnarray}
{\cal L}^{nl}_{{\cal C}\rm H}&=&\frac{ig^2\xi}{4}\epsilon^{3ik}\left[(\phi_0-\phi)^\dag \sigma^k \sigma^j\phi- \phi^\dag \sigma^j \sigma^k(\phi-\phi_0)  \right]\bar{\cal C}^i{\cal C}^j\nonumber \\
&&+\frac{igg'\xi}{4}\epsilon^{3ij}\left[(\phi_0-\phi)^\dag \sigma^j\phi- \phi^\dag \sigma^j(\phi-\phi_0)  \right]\bar{\cal C}^i{\cal C}\, .
\end{eqnarray}
Explicit Feynman rules for antighost-ghost-scalar interactions can be derived from the above Lagrangians by passing from the gauge basis $\{\bar{\cal C}^i, {\cal C}^i, \bar{\cal C}, {\cal C} \}$ to the mass basis  $\{\bar{\cal C}^\pm, {\cal C}^\pm, \bar{\cal C}^Z, {\cal C}^Z,\bar{\cal C}^A, {\cal C}^A \}$, but we do not do it here, since it is not necessary for our calculations.

\begin{table}[htbp]
\centering
\renewcommand{\arraystretch}{1.5}
\begin{tabular}{|c|c|c|c|c|c|c|}
\hline
Vertex Function & SMNLG & SMENLG
\\ \hline
$W^{+\mu}(k)W^{-\nu}(k)$ & $\hat{\Gamma}^{WW}_{\mu \nu}(k)$, \ Eq.(\ref{VFWW1}) & $\hat{\Gamma}^{WW}_{\mu \nu}(k)+\bar{\Gamma}^{WW}_{\mu \nu}(k) $, \ Eqs.(\ref{VFWW1},\ref{VFWW2})
\\ \hline
$G^+_WG^-_W$ & $-i(k_{\phi \phi})_{\alpha \beta}k^\alpha k^\beta $ & $-i\left(k_{\phi \phi}\right)_{\alpha \beta}k^\alpha k^\beta $
\\ \hline
$W^\mp G^\pm(k)$ & $\pm i m_W\left(k_{\phi \phi}-ik_{\phi W}\right)_{\mu \nu}k^\nu$ & $\pm m_W\left(k_{\phi W}\right)_{\mu \nu}k^\nu$
\\ \hline
$A^\mu(k) H$ & $-m_Ws_W(k_{\phi B})_{\mu \nu}k^\nu$ & $-m_Ws_W(k_{\phi B})_{\mu \nu}k^\nu$
\\ \hline
$A^\eta(k_1)W^{+\lambda}(k_2)W^{-\rho}(k_3)$ & $-ie\hat{\Gamma}^{WW\gamma}_{\lambda \rho \eta}(k_1,k_2,k_3)$, \ Eq.(\ref{WWA1}) & $-ie\left(\hat{\Gamma}^{WW\gamma}_{\lambda \rho \eta}(k_1,k_2,k_3)+\bar{\Gamma}^{WW\gamma}_{\lambda \rho \eta}(k_1,k_2,k_3)\right)$, \ Eqs.(\ref{WWA1},\ref{WWA2})
\\ \hline
$A^\alpha A^\beta W^{+\lambda} W^{-\rho}$ & $-ie^2\hat{\Gamma}^{WW\gamma \gamma}_{\alpha \beta \lambda \rho}$, \ Eq.(\ref{WWAA1}) & $-ie^2\left(\hat{\Gamma}^{WW\gamma \gamma}_{\alpha \beta \lambda \rho}+\bar{\Gamma}^{WW\gamma \gamma}_{\alpha \beta \lambda \rho}\right)$, \ Eqs.(\ref{WWAA1},\ref{WWAA2})
\\ \hline
$A^\mu(q)G^+_W(p_1)G^-_W(p_2)$ & $-ie\Gamma^{G_WG_W\gamma}_\mu(p_1,p_2,q)$, \ Eq.(\ref{GGA}) & $-ie\Gamma^{G_WG_W\gamma}_\mu(p_1,p_2,q)$, \ Eq.(\ref{GGA})
\\ \hline
$A^\mu A^\nu G^+_WG^-_W$ & $ie^2\Gamma^{G_WG_W\gamma \gamma}_{\mu \nu}$, \ Eq.(\ref{GGAA}) & $ie^2\Gamma^{G_WG_W\gamma \gamma}_{\mu \nu}$ , \ Eq.(\ref{GGAA})
\\ \hline
$A^\mu W^{\pm \nu}G^{\mp}$ & $em_W\left(ik_{\phi \phi}\mp k_{\phi W}\right)_{\mu \nu}$ & $\mp em_W(k_{\phi W})_{\mu \nu}$
\\ \hline
$A^\mu(q) H H, \ A^\mu(q)G_Z G_Z $ & $-\frac{e}{2}(k_{\phi B})_{\mu \nu}q^\nu$ & $-\frac{e}{2}(k_{\phi B})_{\mu \nu}q^\nu$
\\ \hline
\end{tabular}
\caption{\label{FR} Feynman rules needed to calculate the photon propagator. All momenta are taken incoming. In this Table, the acronym SMNLG refers to a nonlinear gauge-fixing procedure that does not include LV, while SMENLG has been used for a nonlinear gauge-fixing procedure that includes LV. In these gauges, the electromagnetic couplings of pseudo-Goldstone bosons and ghosts coincide.}
\end{table}

\section{One-loop effects on the photon propagator from the Yukawa-sector extension}
\label{FPP}
In this section, we present the results of the one-loop fermionic contribution (charged leptons and quarks) to the photon propagator. This calculation, besides being intricate, has some subtleties. Because of this, we have resorted to the symmetries of the problem and have systematically used the package FeynCalc~\cite{FeynCalc} in all the stages of the calculation. Based on symmetry criteria, one expects, in terms of the irreducible parts of $(k_F)_{\mu \alpha \beta \nu}$, the following type of interactions:
\begin{equation}
\label{GSS}
(\hat{k}_F)_{\mu \alpha \beta \nu}F^{\mu \alpha}F^{\beta \nu}\, , \ \ \ (\tilde{k}_F)_{\mu \alpha \beta \nu}F^{\mu \alpha}F^{\beta \nu}=(k_F)_{\mu \nu}F^{\mu \alpha}F^{\ \nu}_{\alpha}\, , \ \ \  k_FF_{\mu \nu}F^{\mu \nu}\, , \ \ \ (k_F)_{\alpha \beta}\partial^\alpha F_{\mu \nu}\partial^\beta F^{\mu \nu}\, .
\end{equation}
Note that the last term does not have a renormalizable structure, since its mass dimension is 6. Then, as a consequence of the one-loop renormalizability of QEDE~\cite{QEDE}, such a term must be free of ultraviolet divergences. This is also a good criterion to make sure that our results are correct.

\subsection{The calculation}
\begin{figure}
\center
\includegraphics[width=4.0in]{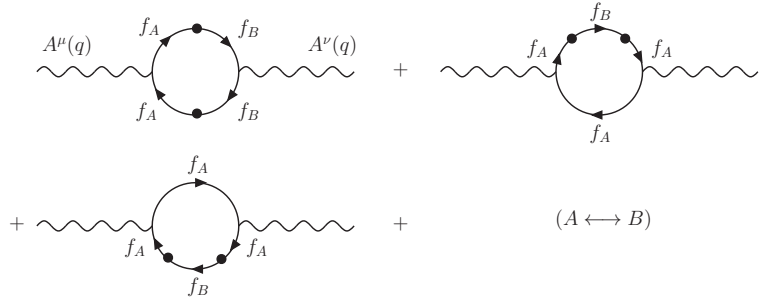}
\caption{\label{FD} Fermion contribution to the photon propagator in the mSME. The black dots denote a flavor-changing insertion $f_Af_B$.}
\end{figure}

Due to gauge invariance, there is no linear contribution, with respect to the antisymmetric tensor $ Y^{AB}_{\alpha \beta} $, to the photon propagator, so we explore the second-order contribution. The Feynman diagrams that can contribute at the second order in $Y^{AB}_{\alpha\beta}$ are shown in Fig.~\ref{FD}. Notice that general flavor-violating effects in each insertion or vertex (black dots) is considered. The corresponding amplitude can be written as follows:
\begin{equation}
\label{FA}
\Pi^f_{\mu \nu}=-m_W^2 s^2_W \sum_{A,B}Q^2_fN_C \int \frac{d^Dk}{(2\pi)^D}\left(\frac{T^{(1)}_{\mu \nu}}{\Delta_A \Delta_B \Delta_{Aq}\Delta_{Bq}}+\frac{T^{(2)}_{\mu \nu}}{\Delta^2_A \Delta_B \Delta_{Aq}}+\frac{T^{(3)}_{\mu \nu}}{\Delta_A\Delta^2_{Aq}\Delta_{Bq}}\right)\, ,
\end{equation}
where $Q_f$ is the charge content of the fermion under consideration, in units of the positron charge, and $N_C$ is the color factor, which is 3 for quarks and 1 for leptons. In addition,
\begin{eqnarray}
T^{(1)}_{\mu \nu}&=&{\rm Tr}\left[\gamma_\mu \Lambda_{Aq}M^{BA*}\Lambda_{Bq}\gamma_\nu \Lambda_BM^{AB}\Lambda_A\right]+(A \leftrightarrow B)\, , \\
T^{(2)}_{\mu \nu}&=&{\rm Tr}\left[\gamma_\mu \Lambda_{Aq}\gamma_\nu \Lambda_AM^{BA*}\Lambda_{B}M^{AB}\Lambda_A\right]+(A \leftrightarrow B)\, , \\
T^{(3)}_{\mu \nu}&=&{\rm Tr}\left[\gamma_\mu \Lambda_{Aq}M^{BA*}\Lambda_{Bq}M^{AB}\Lambda_{Aq}\gamma_\nu \Lambda_A\right]+(A \leftrightarrow B)\, ,
\end{eqnarray}
where
\begin{equation}
M=(V_{\alpha \beta}+A_{\alpha \beta}\gamma_5)\sigma^{\alpha \beta}\, ,
\end{equation}
\begin{eqnarray}
\Lambda_A&=&\pFMSlash{k}+m_A\, , \ \ \ \ \ \ \ \ \Delta_A=k^2-m^2_A\, , \\
\Lambda_B&=&\pFMSlash{k}+m_B\, , \ \ \ \ \ \ \ \ \Delta_B=k^2-m^2_B\, , \\
\Lambda_{Aq}&=&\pFMSlash{k}-\pFMSlash{q}+m_A\, , \ \   \Delta_{Aq}=(k-q)^2-m^2_A\, , \\
\Lambda_{Bq}&=&\pFMSlash{k}-\pFMSlash{q}+m_B\, ,  \ \  \Delta_{Bq}=(k-q)^2-m^2_B \, .
\end{eqnarray}
The amplitude displayed in Eq.~(\ref{FA}) has integrals that are superficially divergent, but such divergences are logarithmic and cancel each other when the partial amplitudes arising from all diagrams are summed together. There are no linear divergences, so the result is finite and unambiguous. The contributions can be divided into those that do not emerge from a trace that involves a $\gamma^5$, which are proportional to tensor products $V_{\alpha\beta}V_{\lambda\rho}$ or $A_{\alpha\beta}A_{\lambda\rho}$, and those which arise from traces containing $\gamma^5$, in whose case the contribution is proportional to the contraction of $V_{\alpha\beta}A_{\lambda\rho}$ with the Levi-Civita tensor. In terms of the ${\rm SO}(1,3)$  irreducible parts of $k_{\alpha \beta \mu \nu}$, the one-loop amplitude can be written as follows:
\begin{equation}
\label{AF}
\Pi^f_{\mu \nu}=\left((\hat{k}^f_F)_{\mu \alpha \beta \nu}+ (\tilde{k}^f_F)_{\mu \alpha \beta \nu}\right)q^\alpha q^\beta+
\left(k^f_F+(k^f_F)_{\alpha \beta}\frac{q^\alpha q^\beta}{q^2}\right)P_{\mu \nu}\, ,
\end{equation}
where
\begin{equation}
P_{\mu \nu}=q^2g_{\mu \nu}-q_\mu q_\nu \, ,
\end{equation}
\begin{eqnarray}
(\hat{k}^f_F)_{\mu \alpha \beta \nu}&=&\frac{i\alpha}{4\pi}\sum_{A,B}Q^2_fN_C \left[f_1(q^2)\left((\hat{k}^{VV}_F)_{\mu \alpha \beta \nu}+(\hat{k}^{AA}_F)_{\mu \alpha \beta \nu}\right)+f_2(q^2)(\hat{k}^{AV}_{F1})_{\mu \alpha \beta \nu}+f_3(q^2)(\hat{k}^{AV}_{F2})_{\mu \alpha \beta \nu}\right]\, ,\\
(\tilde{k}^f_F)_{\mu \alpha \beta \nu}&=&\frac{i\alpha}{4\pi}\sum_{A,B}Q^2_fN_C \left[f_4(q^2)(\tilde{k}^{VV}_F)_{\mu \alpha \beta \nu}+f_5(q^2)(\tilde{k}^{AA}_F)_{\mu \alpha \beta \nu}+f_6(q^2)(\tilde{k}^{AV}_{F1})_{\mu \alpha \beta \nu}+f_7(q^2)(\tilde{k}^{AV}_{F2})_{\mu \alpha \beta \nu}\right]\, ,\\
k^f_{F}&=&\frac{i\alpha}{4\pi}\sum_{A,B}Q^2_fN_C \left[f_8(q^2)k^{VV}_F+f_9(q^2)k^{AA}_F+f_{10}(q^2)k^{AV}_{F}\right]\, , \\
(k^f_F)_{\alpha \beta}&=&\frac{i\alpha}{4\pi}\sum_{A,B}Q^2_fN_C \left[f_3(q^2)\left((k^{VV}_F)_{\alpha \beta}+(k^{AA}_F)_{\alpha \beta}\right)+f_7(q^2)(k^{AV}_{F1})_{\alpha \beta}\right]\, .
\end{eqnarray}
The various Riemann-type tensors that define the above expressions through their irreducible parts (see Eqs.(\ref{RT1}) and (\ref{RT2})) are given by:
\begin{eqnarray}
(k^{VV}_F)_{\mu \alpha \beta \nu}&=& V^{AB}_{\mu \alpha}V^{BA*}_{\beta \nu} \, , \\
(k^{AA}_F)_{\mu \alpha \beta \nu}&=& A^{AB}_{\mu \alpha}A^{BA*}_{\beta \nu} \, ,
\end{eqnarray}
\begin{eqnarray}
(k^{AV}_{F1})_{\mu \alpha \beta \nu}&=&A^{AB}_{\mu \alpha}\tilde{V}^{BA*}_{\beta \nu}+A^{AB}_{\nu \beta}\tilde{V}^{BA*}_{\alpha \mu}\, ,\\
(k^{AV}_{F2})_{\mu \alpha \beta \nu}&=&\left(A_\mu^{AB\ \lambda} V_{\alpha}^{BA*\ \rho}-A_\alpha^{AB\ \lambda} V_{\mu}^{BA* \ \rho}\right)\epsilon_{\lambda \rho \nu \beta}\nonumber \\
&&- \left(A_\beta^{AB\ \lambda} V_{\nu}^{BA*\ \rho}-A_\nu^{AB\ \lambda} V_{\beta}^{BA*\ \rho}\right)\epsilon_{\lambda \rho \mu \alpha}\, ,
\end{eqnarray}
where $\tilde{V}_{\alpha \beta}=(1/2)\epsilon_{\alpha \beta \lambda \rho}V^{\lambda \rho}$. Notice that $k^{AV}_{F1}=4k^{AB}_{F2}\equiv k^{AV}_F$. The dimensionless form factors $f_i(q^2)$, which are listed in Appendix \ref{FF}, are all free of ultraviolet divergences. Note that the last term of Eq.(\ref{AF}) is given only in terms of symmetric Ricci-type tensors $(k^{VV}_F)_{\alpha \beta}$, etc., and corresponds to a dimension-six effect, which is thus not present at the tree level.

\subsection{Discussion}
Including the one-loop contribution from the Yukawa-sector extension given in Eq.(\ref{AF}), the Lagrangian for the pure-photon sector, Eq.~(\ref{PL}), can be written as
\begin{eqnarray}
\label{LFL}
{\cal L}^{f\, \text{1-loop}}_{\rm photon}&=&-\frac{1}{4}\left(1+k^f_F\right)F_{\mu \nu}F^{\mu \nu}-\frac{1}{4}\left((\hat{k}_F)_{\mu \alpha \beta \nu}-(\hat{k}^f_F)_{\mu \alpha \beta \nu}\right)F^{\mu \alpha}F^{\beta \nu}\nonumber \\
&&-\frac{1}{4}\left((k_F)_{\mu \nu}-(k^f_F)_{\mu \nu}\right)F^{\mu \alpha}F^{\  \nu}_\alpha+\frac{(k^f_F)_{\alpha \beta}}{4v^2}\,\partial^\alpha F_{\mu \nu}\partial^\beta F^{\mu \nu}\, ,
\end{eqnarray}
where we have used the relation between the tensors $(\tilde{k}_F)_{\mu \alpha \beta \nu}$ and $(k_F)_{\alpha \beta}$, given by Eq.(\ref{RT2}), so that the Lagrangian (\ref{LFL}) is expressed only in terms of the irreducible parts of $k_{\mu \alpha \beta \nu}$. Moreover, we have not included the CS term, as it does not receive any contribution from the extended Yukawa sector. With the exception of the dimension-6 term, all contributions from the Yukawa sector are unobservable at this level, as they can be removed through a finite renormalization of the field, the electric charge, and the tensors $(\hat{k}_F)_{\mu \alpha \beta \nu}$ and $(k_F)_{\alpha \beta}$. Concretely, this is achieved through the following redefinitions:
\begin{eqnarray}
A_\mu &\to& \left(1+k^f_F\right)^{-\frac{1}{2}}A_\mu \, , \\
e&\to & \left(1+k^f_F\right)^{\frac{1}{2}}e\, , \\
(\hat{k}_F)_{\mu \alpha \beta \nu} &\to&  \left(1+k^f_F\right)(\hat{k}_F)_{\mu \alpha \beta \nu}+(\hat{k}^f_F)_{\mu \alpha \beta \nu}\, , \\
(k_F)_{ \alpha \beta} &\to&  \left(1+k^f_F\right)(k_F)_{\alpha \beta }+(k^f_F)_{\alpha \beta}\, ,
\end{eqnarray}
which leads to the Lagrangian
\begin{equation}
{\cal L}^{f\, \text{1-loop}}_{\rm photon}=-\frac{1}{4}F_{\mu \nu}F^{\mu \nu}-\frac{1}{4}(k_F)_{\mu \alpha \beta \nu}F^{\mu \alpha}F^{\beta \nu}+\frac{(k^f_F)_{\alpha \beta}}{4v^2}\left(1+k^f_F\right)^{-1}\,\partial^\alpha F_{\mu \nu}\partial^\beta F^{\mu \nu}\, .
\end{equation}
This result shows that, at one loop, the most general extension of renormalizable type of the Yukawa sector does not contribute to photon birefringence in vacuum. At this level, the only contribution emerges via a dimension-6 term proportional to the Ricci-type tensor $(k_F)_{\alpha \beta}$, which does not contribute to birefringence.

\section{One-loop effects on the photon propagator from the Higgs-sector extension}
\label{HS} In this section, we study the impact from the Higgs-sector extension on the photon propagator. Contributions can arise from both the symmetric constant tensor $(k_{\phi \phi})_{\alpha \beta}$ and the antisymmetric constant tensors $(k_{\phi W})_{\alpha \beta}, (k_{\phi B})_{\alpha \beta}$. We discuss each type of contribution separately.

\subsection{Antisymmetric contribution}
\begin{figure}
\center
\includegraphics[width=5.0in]{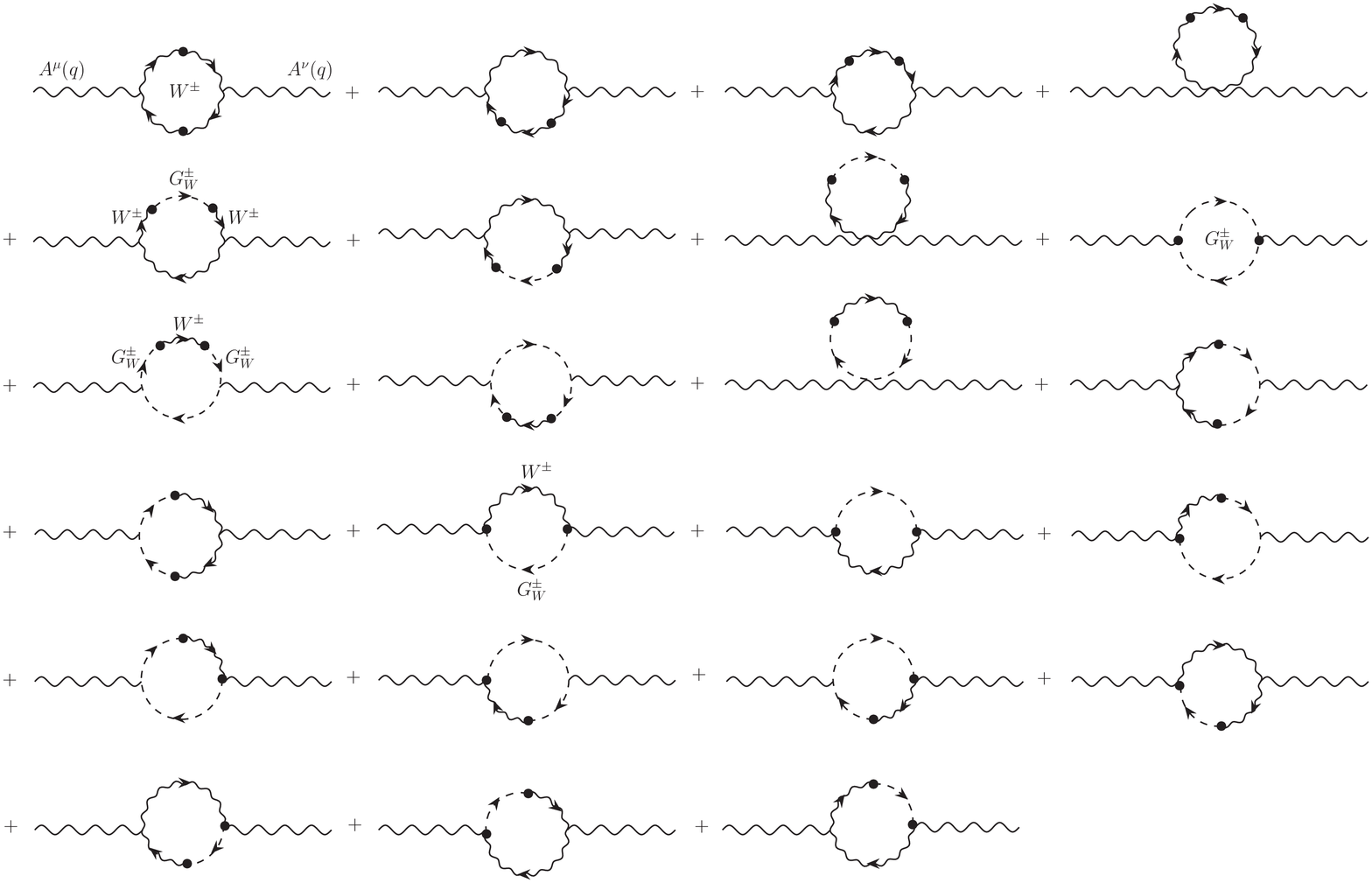}
\caption{\label{W1} Contributions of the antisymmetric tensors $k^A_{\phi \phi}$, $k_{\phi W}$ and $k_{\phi B}$ to the photon propagator in a nonlinear gauge. Black dots denote an insertion of a bilinear coupling or a nonstandard trilinear vertex.}
\end{figure}

Due to gauge invariance, the contribution from the antisymmetric background tensors to the photon propagator is, necessarily, of second order in such  tensors. We carry out the calculation in a nonlinear gauge, which, as already discussed in Sec.~\ref{GFP}, allows us to remove the standard couplings $WG_W$ and $WG_W\gamma$. It should be recalled that these vertices are not removed from the antisymmetric part of the Higgs sector. The Feynman rules needed for the calculation are shown in Table~\ref{FR}. In the Feynman-'t Hooft version of this gauge, the contribution from the set $\{(k_{\phi W})_{\alpha \beta}, (k_{\phi B})_{\alpha \beta}\}$ of tensors is given by the diagrams shown in Fig.~\ref{W1}.

\begin{figure}
\center
\includegraphics[width=3.5in]{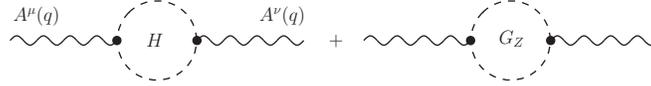}
\caption{\label{HGz} Higgs and pseudo-Goldstone boson $G_Z$ contribution to the photon propagator. Black dots denote nonstandard trilinear vertices proportional to the $k_{\phi B}$ constant tensor.}
\end{figure}

In this case, we organize the contributions in terms of four Riemann-type tensors given by
\begin{eqnarray}
(k^{WW}_{F})_{\mu \alpha \beta \nu}&=&(k_{\phi W})_{\mu \alpha}(k_{\phi W})_{\beta \nu}\, , \\
(k^{WB}_{F})_{\mu \alpha \beta \nu}&=&(k_{\phi W})_{\mu \alpha}(k_{\phi B})_{\beta \nu}+(k_{\phi B})_{\mu \alpha}(k_{\phi W})_{\beta \nu}\, ,\\
(k^{BB}_{F})_{\mu \alpha \beta \nu}&=&(k_{\phi B})_{\mu \alpha}(k_{\phi B})_{\beta \nu}\, .
\end{eqnarray}
The contribution from $(k^{BB}_{F})_{\mu \alpha \beta \nu}$ is given through diagrams shown in Fig.~\ref{HGz}. As in the fermionic case, all gauge structures shown in Eq.~(\ref{GSS}) are generated.

In terms of the irreducible parts of the above Riemann-type tensors, the amplitude can be written as follows:
\begin{eqnarray}
\label{HA}
\Pi^{(H,A)}_{\mu \nu}&=&\Big\{g_1(q^2)(\hat{k}^{WW}_F)_{\mu \alpha \beta \nu}+g_2(q^2)(\tilde{k}^{WW}_F)_{\mu \alpha \beta \nu}+
g_3(q^2)\left[(\hat{k}^{BB}_F)_{\mu \alpha \beta \nu}+(\tilde{k}^{BB}_F)_{\mu \alpha \beta \nu}\right] \nonumber \\
&&+g_4(q^2)\left[(\hat{k}^{WB}_F)_{\mu \alpha \beta \nu}+(\tilde{k}^{WB}_F)_{\mu \alpha \beta \nu}\right]\Big\}q^\alpha q^\beta \nonumber \\
&&+\left[g_5(q^2)k^{WW}_F-\frac{1}{6}g_3(q^2)k^{BB}_F-\frac{1}{6}g_4(q^2)k^{WB}_F +g_6(q^2)(k^{WW}_F)_{\alpha \beta}\left(\frac{q^\alpha q^\beta}{q^2}\right) \right]P_{\mu \nu}\, ,
\end{eqnarray}
where the form factors $g_i(q^2)$, with $i=1,2,3,4,5$, are all divergent. On the other hand, the form factor associated with the dimension-6 term, $g_6(q^2)$, is free of ultraviolet divergences, which is in accordance with the fact that mQEDE is renormalizable. These functions are listed in Appendix~\ref{BF}.

\subsubsection{Discussion}
As it can be seen from Eq.~(\ref{HA}), all the antisymmetric parts of the extended Higgs sector induce divergent contributions to the Weyl-type tensors, so they are severely restricted by vacuum birefringence. Notice that, as we mentioned in the Introduction, all the scalar coefficients $k^{WW}_F$,  $k^{BB}_F$, and $k^{WB}_F$ are unobservable because they can be absorbed through the renormalization of the electromagnetic field. Bounds on the components of the birefringent parts, $(\hat{k}^{WW}_F)_{\mu \alpha \beta \nu}$, $(\hat{k}^{BB}_F)_{\mu \alpha \beta \nu}$, and $(\hat{k}^{WB}_F)_{\mu \alpha \beta \nu}$ have already been derived in Ref.~\cite{Sher}. The authors of that reference assume that no cancelations between the various types of contributions occur and then they constrain each one at a time. Focusing only on the divergent part of the amplitudes and then imposing a cut off on them, they derive bounds of $\sim3\times 10^{-16}$ for the components of~\footnote{The authors of this reference also derived bounds for the componentes of $|(k^A_{\phi \phi})_{\alpha \beta}|$, but as we seen in Sec.~\ref{FW} this tensor does not really exist.} $|(k_{\phi W})_{\alpha \beta}|$, and $|(k_{\phi B})_{\alpha \beta}|$. Of course, since the mQEDE is renormalizable, in principle, it may be possible to introduce a renormalization scheme that allows us to derive bounds on these tensor objects, but its implementation may be quite complicated in practice. For comparison purposes, here we follow a simpler approach, suggested in Ref.~\cite{SME}. It consists in considering the one-loop contribution as a radiative correction to the renormalized parameter, which is assumed to be of the same order of magnitude, that is, we assume that $\alpha |(k_{\phi W})_{\alpha \beta}|^2\sim (\tilde{\kappa}_{e^+},\tilde{\kappa}_{o^-})<10^{-32}$, etc., which leads to bounds $\sim10^{-15}$ for $|(k_{\phi W})_{\alpha \beta}|$ and $|(k_{\phi B})_{\alpha \beta}|$. This shows that the approach followed in Ref.~\cite{Sher} coincides essentially with the one suggested in Ref.~\cite{SME}.

\subsection{Symmetric contribution}
This contribution is given only by the symmetric $(k_{\phi \phi})_{\alpha \beta}$ tensor. In Sec.~\ref{GFP}, we saw that the $(k_{\phi \phi})_{\alpha \beta}$ tensor may or may not be included in a nonlinear gauge-fixing procedure. For purposes that will be clear later, we will carry out the calculation with both types of nonlinear gauges.

\subsubsection{SM nonlinear gauge}
\begin{figure}
\center
\includegraphics[width=4.5in]{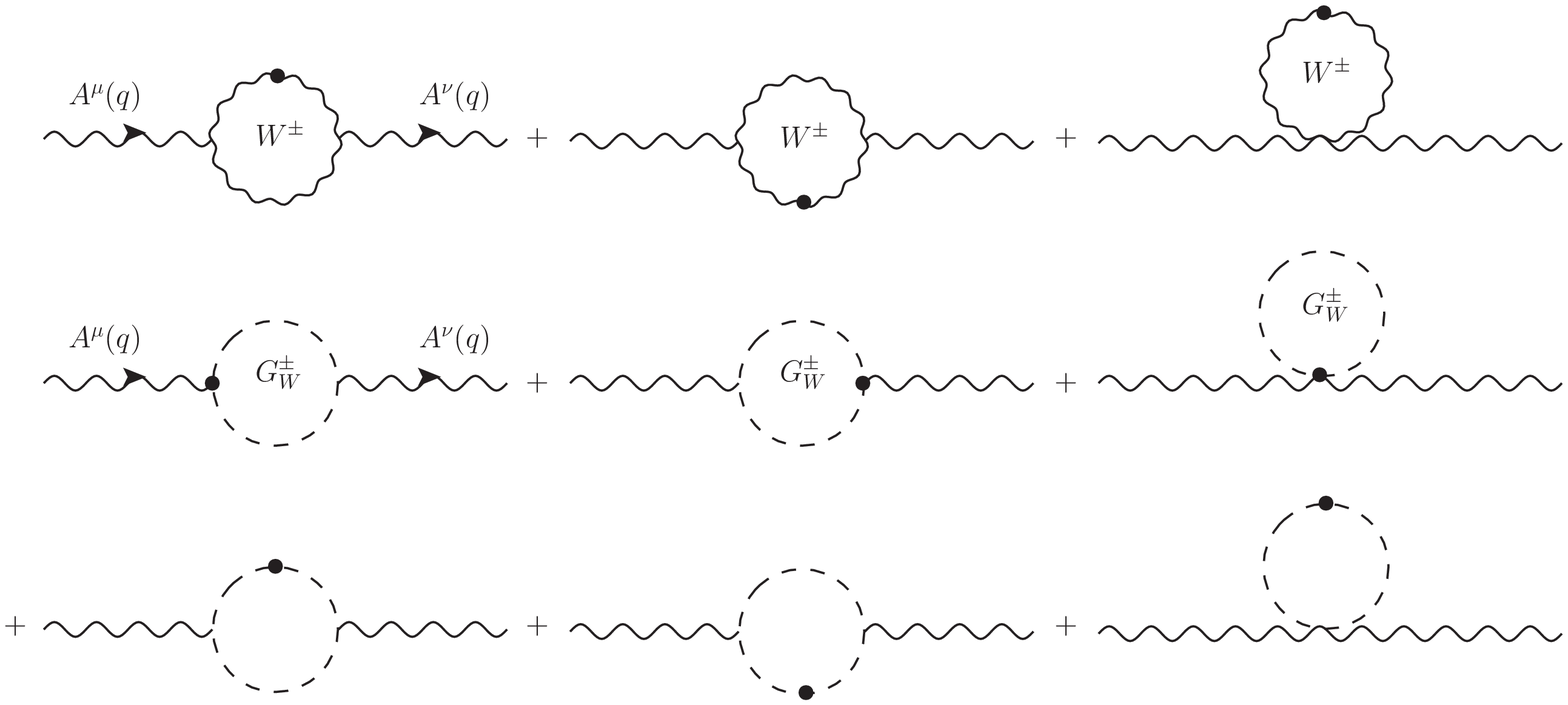}
\caption{\label{W2} Diagrams contributing to the non-birefringent part of $(k_F)_{\mu \alpha \beta \nu}$ in a nonlinear gauge which does not incorporates LV. There is no ghost contribution.}
\end{figure}

In this case, $(k_{\phi \phi})_{\alpha \beta}$-linear gauge-invariant contributions are generated through diagrams shown in Fig.~\ref{W2}. The Feynman rules needed for the calculation are shown in the second column of Table~\ref{FR}. It turns out that the diagrams in which only the $W$ gauge boson circulates (first row of Fig.~\ref{W2}) lead, by themselves, to a result that is finite and gauge invariant. On the other hand, the diagrams in which only the pseudo-Goldstone boson $G_W$ circulates also lead to a gauge-invariant result, which, however, is divergent. The corresponding amplitude can be written as follows:
\begin{equation}
\label{HSSM}
\hat{\Pi}^{(H,S)}_{\mu \nu}=\left[\hat{g}_1(q^2)k_{\phi \phi}+\hat{g}_3(q^2)(k_{\phi \phi})_{\alpha \beta}\left(\frac{q^\alpha q^\beta}{q^2}\right)\right]P_{\mu \nu}
+\hat{g}_2(q^2)(\tilde{k}_{\phi \phi})_{\mu \alpha \beta \nu}\, q^\alpha q^\beta \, ,
\end{equation}
where the $(\tilde{k}_{\phi \phi})_{\mu \alpha \beta \nu}$ tensor is defined as products of the metric tensor and the $(k_{\phi \phi})_{\alpha \beta}$ tensor through Eq.~(\ref{RT2}). In addition, $k_{\phi \phi}=g^{\alpha \beta}(k_{\phi \phi})_{\alpha \beta}$ and the form factors $\hat{g}_i(q^2)$ are listed in Appendix~\ref{BF}. The $\hat{g}_1(q^2)$ and $\hat{g}_2(q^2)$ form factors are divergent, but $\hat{g}_3(q^2)$ is finite, in accordance with renormalizaion theory, as it characterizes a dimension-6 interaction.

\subsubsection{SME nonlinear gauge}
\begin{figure}
\center
\includegraphics[width=4.5in]{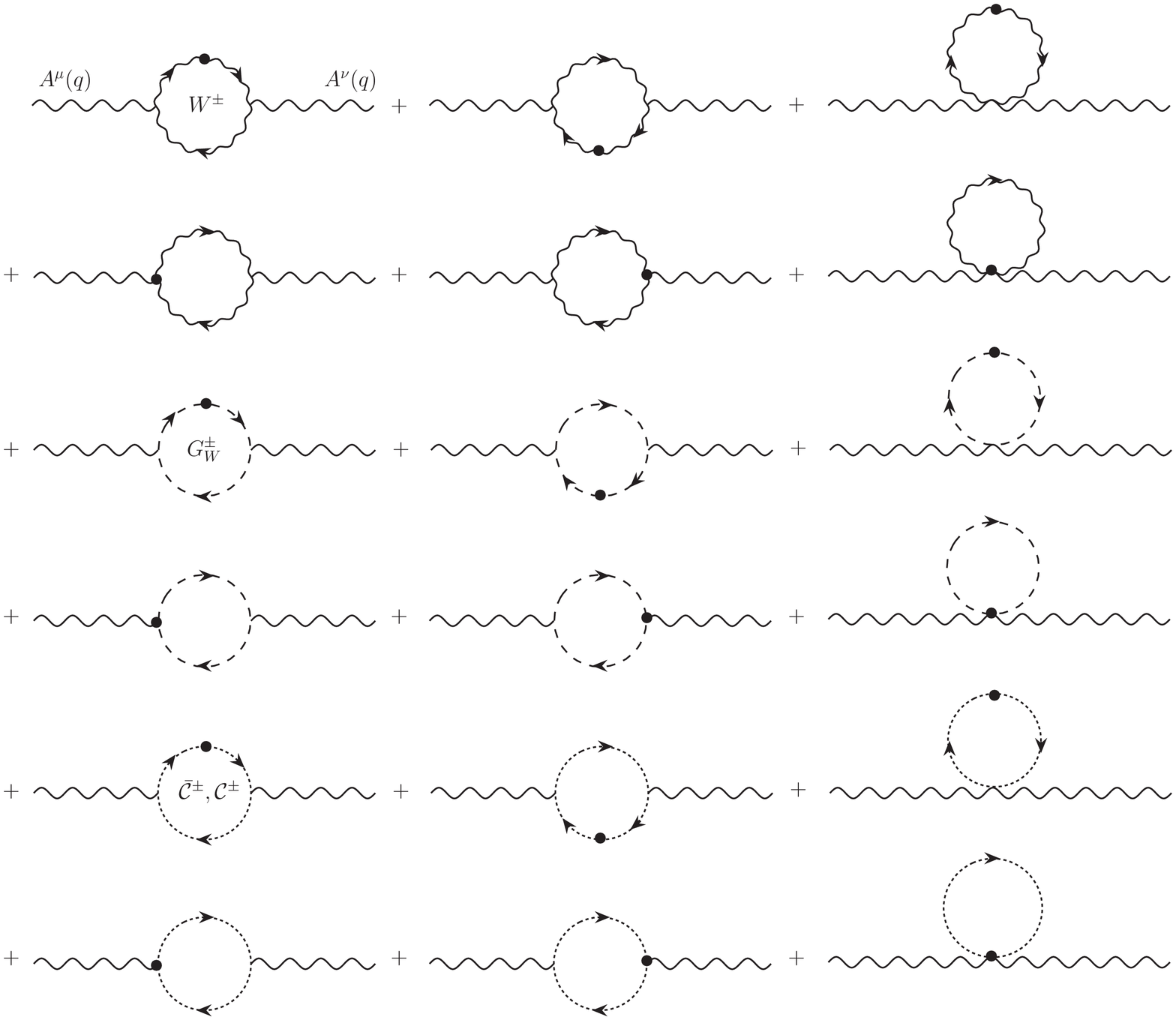}
\caption{\label{W3} Diagrams contributing to the non-birefringent part of $(k_F)_{\mu \alpha \beta \nu}$ in a nonlinear gauge which incorporates  LV. Ghost contribution is represented by diagrams with dotted lines.}
\end{figure}

We now turn to calculate the  $(k_{\phi \phi})_{\alpha \beta}$ contribution to the photon propagator in the context of the more general nonlinear gauge discussed in Sec.~\ref{GFP}, which includes this tensor in the definition of the gauge-fixing functions. It should be remembered that this gauge allows us to remove, among others, the unphysical couplings $WG_W$ and $\gamma WG_W$ both in the SM model part and in the mSME (only those that are proportional to the $(k_{\phi \phi})_{\alpha \beta}$ tensor). In this case, the Feynman rules needed for the calculation are shown in the third column of Table~\ref{FR}. The contribution is given through the diagrams shown in Fig.~\ref{W3}. In this gauge, each type of particle circulating in the loops leads by itself to a gauge invariant result and also to a finite contribution to the form factor associated with the dimension-6 interaction. Due to the symmetric structure of the ghost sector under the $U_Q(1)$ gauge group, its contribution is exactly minus twice the one of the pseudo-Goldstone boson. The corresponding amplitude can be written as follows:

\begin{equation}
\label{HSSME}
\bar{\Pi}^{(H,S)}_{\mu \nu}=\left[\bar{g}_1(q^2)k_{\phi \phi}+\bar{g}_3(q^2)(k_{\phi \phi})_{\alpha \beta}\left(\frac{q^\alpha q^\beta}{q^2}\right)\right]P_{\mu \nu}
+\bar{g}_2(q^2)(\tilde{k}_{\phi \phi})_{\mu \alpha \beta \nu}\, q^\alpha q^\beta \, ,
\end{equation}
with the the form factors $\bar{g}_i(q^2)$ listed in Appendix~\ref{BF}.

\subsubsection{Discussion}
\label{CNB}
Above, we have calculated the contribution from the symmetric part of the Higgs sector to the photon propagator in two gauge-fixing procedures that are essentially different. It is worth highlighting some important aspects of these results. In first place is the fact that the poles of the divergent form factors are the same in both gauge-fixing procedures, that is, ${\rm Div}(\hat{g}_1)={\rm Div}(\bar{g}_1)=\frac{i\alpha}{4\pi}\left(\frac{B_0(2)}{6}\right)=\frac{i\alpha}{24\pi\epsilon}+\cdots$ and ${\rm Div}(\hat{g}_2)={\rm Div}(\bar{g}_2)=\frac{i\alpha}{4\pi}\left(\frac{-2B_0(2)}{3}\right)=\frac{-i\alpha}{6\pi\epsilon}+\cdots$ (see Appendix~\ref{BF}), where the spacetime dimension of dimensional regularization is $D=4-2\epsilon$. Since the beta functions are determined essentially by the coefficient of the pole, this result reflects the fact that the corresponding beta function is gauge-independent. Another interesting result is that $\bar{g}_3(q^2)=\hat{g}_3(q^2)$, that is, the finite form factors associated with the dimension-6 coupling coincide in both gauge-fixing schemes, which suggests gauge independence of this term. However notice that, to be sure of this, one should also show that the result does not depend on the gauge parameter $\xi$.

From Eqs.~(\ref{HSSM}) and (\ref{HSSME}), we see that the symmetric part of the extended Higgs sector does not contribute to birefringence, so other means must be used to bound the components of the symmetric tensor $(k_{\phi \phi})_{\alpha \beta}$. Limits derived from the properties of beta decay are available in the literature. In~\cite{HWW} a bound at a level of $10^{-3}$ was obtained, which was improved in the same context at a level of $10^{-4}$ in~\cite{OBS}. Here, we explore the possibility of constraining the components of $(k_{\phi \phi})_{\alpha \beta}$ directly from bounds derived for the matrices $(\tilde{\kappa}_{e^-},\tilde{\kappa}_{o^+})$, which, as commented in the Introduction, were obtained~\cite{BNBR} from the Laser Interferometer Gravitational-Wave Observatory. To our knowledge, these are the best bounds available so  far~\cite{DT}. The  Ricci-type tensor $(k_F)_{\alpha \beta}$ defines the non-birefringent tensor $(\tilde{k}_F)_{\mu \alpha \beta \nu}$ through Eq.~(\ref{RT2}), which in turn is parametrized by the five components of the symmetric traceless matrix $\tilde{\kappa}_{e^-}$, by the three components of the antisymmetric matrix $\tilde{\kappa}_{o^+}$, and by the parameter $\tilde{\kappa}_{tr}$ (see, for instance, Ref.~\cite{Casana}). Such a connection is given by
\begin{eqnarray}
\label{R1}
(\tilde{\kappa}_{e^-})^{ij}&=&\delta^{ij}(k_F)^{00}-(k_F)^{ij}-\delta^{ij}\tilde{\kappa}_{tr}\, , \\
\label{R2}
(\tilde{\kappa}_{0^+})^{ij}&=&-\epsilon^{ijk}(k_F)^{0k}\, .
\end{eqnarray}
The bounds derived in Ref.~\cite{BNBR} on the components  of the $(\tilde{\kappa}_{e^-},\tilde{\kappa}_{o^+})$ matrices can directly be translated into the components of the Ricci-type  $(k_F)_{\alpha \beta}$ tensor via Eqs.~(\ref{R1}) and (\ref{R2}). The symmetric matrix $\tilde{\kappa}_{e^-}$ imposes bounds on the space-space components of the $(k_F)_{\alpha \beta}$ tensor given by:
\begin{eqnarray}
|(\tilde{\kappa}_{e^-})^{XY}|&=&|(k_F)^{12}|<2.7\times 10^{-22}\, \\
|(\tilde{\kappa}_{e^-})^{XZ}|&=&|(k_F)^{13}|<2.1\times 10^{-20}\, \\
|(\tilde{\kappa}_{e^-})^{YZ}|&=&|(k_F)^{23}|<2.1\times 10^{-20}\, \\
|(\tilde{\kappa}_{e^-})^{XX}-(\tilde{\kappa}_{e^-})^{YY}|&=&|(k_F)^{22}-(k_F)^{11}|<5.5\times 10^{-22}\, .
\end{eqnarray}
Note that the last bound is valid only if LV is not isotropic. On the other hand, the bounds on the antisymmetric matrix $\tilde{\kappa}_{o^+}$ translate into the space-time components of $(k_F)_{\alpha \beta}$ as follows:
\begin{eqnarray}
|(\tilde{\kappa}_{o^+})^{XY}|&=&|(k_F)^{03}|<6.6\times 10^{-15}\, ,\\
|(\tilde{\kappa}_{o^+})^{XZ}|&=&|(k_F)^{02}|<5.7\times 10^{-15}\, ,\\
|(\tilde{\kappa}_{o^+})^{YZ}|&=&|(k_F)^{01}|<5.2\times 10^{-15}\, .
\end{eqnarray}

As in the birefringent case, we assume that $\alpha (k_{\phi \phi})_{\alpha \beta}$ constitutes a radiative correction to the renormalized tensor $(k_F)_{\alpha \beta}$ and that it is of the same order of magnitude. Then, using the above constraints, we obtain bounds of the following orders of magnitude:
\begin{eqnarray}
|(k_{\phi \phi})^{12}|&<& 10^{-20}\, \\
|(k_{\phi \phi})^{13}|&<& 10^{-18}\, \\
|(k_{\phi \phi})^{23}|&<& 10^{-18}\, \\
|(k_{\phi \phi})^{22}-(k^S_{\phi \phi})^{11}|&<& 10^{-19}\, .
\end{eqnarray}
As far as the space-time components are concerned, we have
\begin{eqnarray}
|(k_{\phi \phi})^{03}|&<& 10^{-12}\, ,\\
|(k_{\phi \phi})^{02}|&<& 10^{-12}\, ,\\
|(k_{\phi \phi})^{01}|&<& 10^{-12}\, .
\end{eqnarray}
From the above results, we see that our bounds obtained directly from limits on the non-birefringent tensor $(\tilde{k}_F)_{\mu \alpha \beta \nu}$ are much stronger than those obtained in~\cite{OBS} from properties of $\beta$ decay.

\section{One-loop effects on the photon propagator from the gauge-sector extension}
\label{GS}
In this section, we study the contribution from the Yang-Mills-sector extension to the photon propagator, which is characterized by the Riemann-type tensor $(k_W)_{\mu \alpha \beta \nu}$. Since the nonlinear gauge-fixing functions introduced in Sec.~\ref{GFP} do not involve this tensor, the corresponding contribution is given only by the $WW$ insertion and the $WW\gamma$ vertex in Feynman diagrams in which the $W$ gauge boson circulates, such as those shown in the first and second rows of Fig.~\ref{W3}. The Feynman rules needed to perform the calculation are given in Table~\ref{FR}.

A direct calculation yields results proportional to the Riemann-type $(k_W)_{\mu \alpha \beta \nu}$ tensor, the Ricci-type $(k_W)_{\alpha \beta}=g^{\mu \nu}(k_W)_{\mu \alpha \nu \beta}$ tensor, and the $k_W=g^{\mu \beta}g^{\alpha \nu}(k_W)_{\mu \alpha \beta \nu}$ scalar. The dimension-6 effect that we have been finding throughout the calculation is generated as well. After using Eqs.~(\ref{RT1}) and (\ref{RT2}) to express the amplitude in terms of the ${\rm SO}(1,3)$ irreducible parts of $(k_W)_{\mu \alpha \beta \nu}$, we have
\begin{equation}
\label{GSC}
\Pi^W_{\mu \nu}=\left[\tilde{g}_1(q^2)(\hat{k}_W)_{\mu \alpha \beta \nu}+\tilde{g}_2(q^2)(\tilde{k}_W)_{\mu \alpha \beta \nu} \right]q^\alpha q^\beta+
\left[\tilde{g}_3(q^2)k_W+\tilde{g}_4(q^2)(k_W)_{\alpha \beta}\left(\frac{q^\alpha q^\beta}{q^2}\right) \right]P_{\mu \nu} \, ,
\end{equation}
where the form factors $\tilde{g}_i(q^2)$ are given in Appendix~\ref{BF}. As in previous cases, the form factor $\tilde{g}_4(q^2)$ associated with the dimension-6 contribution is free of ultraviolet divergences, but all other form factors are divergent.

\subsection{Discussion}
As commented in the Introduction, the birefringent part of the $(k_F)_{\mu \alpha \beta \nu}$ tensor characterized by the Wely-type $(\hat{k}_F)_{\mu \alpha \beta \nu}$ tensor is constrained to be less than $10^{-32}$~\cite{BKF}, which is obtained at the tree level from the Lagrangian~(\ref{PL}). Since $(k_F)_{\mu \alpha \beta \nu}$ is related to $(k_W)_{\mu \alpha \beta \nu}$ and $(k_B)_{\mu \alpha \beta \nu}$ by gauge invariance through the relation $k_F=s^2_Wk_W+c^2_Wk_B$ (see Sec.~\ref{FW}), one would expect a limit for $(\hat{k}_W)_{\mu \alpha \beta \nu}$ of the same order of magnitude as that of $(\hat{k}_F)_{\mu \alpha \beta \nu}$. Alternatively, we can assume, as we have been doing so far, that $\alpha (\hat{k}_W)_{\mu \alpha \beta \nu}$ is the radiative correction to the renormalized $(\hat{k}_F)_{\mu \alpha \beta \nu}$ tensor, which is of the same order of magnitude, that is, the components of $(\hat{k}_W)_{\mu \alpha \beta \nu}$ would be restricted to be lower than $10^{-30}$. As far as the non-birefringent part $(k_W)_{\alpha \beta}$ is concerned, we can apply the same approach followed in Sec.~\ref{CNB} to obtain limits similar to those derived for $(k_{\phi \phi})_{ \alpha \beta}$.

\section{The beta function}
\label{bF}
In Secs.~\ref{FPP}, \ref{HS}, and \ref{GS} we presented the one-loop contribution generated by the Yukawa, Higgs, and gauge sectors of the mSME on the photon propagator. These results, together with the one given in Ref.~\cite{QEDE} in the context of the mQEDE, complete the one-loop radiative correction to the photon propagator in the context of the mSME. We would like to conclude this study by presenting the beta function associated with the $(k_F)_{\mu \alpha \beta \nu}$ tensor. The corresponding contribution from mQEDE has been already calculated in~\cite{QEDE}. It was found in this reference that, apart of the standard QED contribution, the only Lorentz-violating contribution arises from the $\frac{i}{2}c^{\mu \nu}\bar{\psi}\gamma_\mu \overleftrightarrow{D_\nu}\psi$ term, with $c^{\mu \nu}$ a real constant tensor; only the symmetric part of this tensor contributes.

Instead of the  Riemann-type tensor $(k_F)_{\mu \alpha \beta \nu}$, we consider its ${\rm SO}(1,3)$ irreducible parts, namely, the Weyl-type tensor $(\hat{k}_F)_{\mu \alpha \beta \nu}$ and the Ricci-type tensor  $(k_F)_{ \alpha \beta}$, as the most fundamental objects. The reason for this is that, as shown in previous sections, radiative corrections impact these objects differently and are not always generated simultaneously.  So, the bare Lagrangian for the pure photon sector can be written as follows:
\begin{equation}
\label{BL}
{\cal L}_B=-\frac{1}{4}F_{B\mu \nu}F^{\mu \nu}_B-\frac{1}{4}(\hat{k}_F)_{B\mu \alpha \beta \nu}F^{\mu \alpha}_B F^{\beta \nu}_B-\frac{1}{4}(k_F)_{B\alpha \beta}F^{\alpha \gamma}_BF_{B\gamma}^{\, \, \beta}\, ,
\end{equation}
where the CS term $(k_{AF})^\kappa$ has not be included. Also, it has been assumed a double traceless Riemann-type tensor $(k_F)_{\mu \alpha \beta \nu}$, so the scalar $k_F$ is zero. Following Ref.~\cite{QEDE}, to renormalize this part of the theory, we redefine the bare electromagnetic field, the Weyl-type $(\hat{k}_F)_{\mu \alpha \beta \nu}$ tensor, and the Ricci-type $(k_F)_{\alpha \beta}$ tensor in terms of renormalized ones as follows:
\begin{equation}
A_{B\mu}=\sqrt{Z_A}A_\mu\, , \ \ \  (\hat{k}_F)_{B\mu \alpha \beta \nu}=(Z_{\hat{k}_F})_{\mu \alpha \beta \nu}^{\,\, \, \, \, \, \, \, \, \, \, \, \, \gamma \delta \lambda \rho}(\hat{k}_F)_{ \gamma \delta \lambda \rho}\, , \ \ \ (k_F)_{B\alpha \beta}=(Z_{k_F})_{\alpha \beta}^{\, \, \, \, \, \, \lambda \rho}(k_F)_{\lambda \rho}\, ,
\end{equation}
so the bare Lagrangian (\ref{BL}) splitting into the renormalized Lagrangian and the counterterm,
\begin{eqnarray}
{\cal L}_B&=&-\frac{1}{4}F_{\mu \nu}F^{\mu \nu}-\frac{1}{4}(\hat{k}_F)_{\mu \alpha \beta \nu}F^{\mu \alpha} F^{\beta \nu}-\frac{1}{4}(k_F)_{\alpha \beta}F^{\alpha \gamma}F_\gamma^{\, \, \beta}\nonumber \\
&&-\frac{1}{4}(Z_A-1)F_{\mu \nu}F^{\mu \nu}-\frac{1}{4}\left[Z_A(Z_{\hat{k}_F})_{\mu \alpha \beta \nu}^{\,\, \, \, \, \, \, \, \, \, \, \, \, \gamma \delta \lambda \rho}(\hat{k}_F)_{ \gamma \delta \lambda \rho}- (\hat{k}_F)_{\mu \alpha \beta \nu}\right]F^{\mu \alpha} F^{\beta \nu}\nonumber \\
&&-\frac{1}{4}\left[Z_A(Z_{k_F})_{\alpha \beta}^{\, \, \, \, \, \, \lambda \rho}(k_F)_{\lambda \rho}-(k_F)_{\alpha \beta} \right]F^{\alpha \gamma}F_{\gamma}^{\, \, \beta}\, ,
\end{eqnarray}
In the $\overline{MS}$-scheme, the renormalization factor $Z_A$ is exclusively determined by the usual SM contribution,
\begin{equation}
Z_A=1-N\frac{\alpha}{3\pi \, \epsilon}\, , \ \ \ N=\sum_{f=l_A,q_A}N_CQ_f^2-\frac{21}{4}\, ,
\end{equation}
where the $N$ factor quantifies the contributions from charged leptons, quarks, and the $W$ gauge boson. In addition, $\alpha=\frac{e^2}{4\pi}$ and $\epsilon=2-\frac{D}{2}$ in dimensional regularization. On the other hand, in the context of the mQEDE, it was found in Ref.~\cite{QEDE} that the $\frac{i}{2}c^{\mu \nu}\bar{\psi}\gamma_\mu \overleftrightarrow{D_\nu}\psi$ term only contributes to the Ricci-type tensor or, equivalently, to the non-birefringent tensor $(\tilde{k}_F)_{\mu \alpha \beta \nu}$ (see Eqs.~(\ref{RT2}) and (\ref{GSS})). In the case of the Higgs sector extension, we can see from Eq.(\ref{HA}) that the antisymmetric part contributes to both the Weyl-type tensor and the Ricci-type tensor. However, in this case, the divergent part of the corresponding form factors coincide, as can be easily verified from expressions for the $g_1(q^2)$ and $g_2(q^2)$ functions given in Appendix~\ref{BF}. With regard to the symmetric part of this sector, we can see from (\ref{HSSM}) or (\ref{HSSME}) that it contributes only to the Ricci-type tensor. In the gauge sector extension, there are contributions to both the Weyl-type tensor and the Ricci-type tensor, as it can be appreciated from Eq.~(\ref{GSC}). However, in this case, the poles of the corresponding form factors $\tilde{g}_1(q^2)$ and $\tilde{g}_2(q^2)$ do not coincide (see Appendix~\ref{BF}). Then, at order $\alpha$, we can write the corresponding renormalization factors as follows:
\begin{equation}
\label{Z1}
(Z_{k_F})^{\alpha \beta}_{\, \, \, \, \, \, \, \lambda \rho}(k_F)_{\lambda \rho}=(k_F)^{\alpha \beta} +N\frac{\alpha}{3\pi \epsilon}\left[(k_F)^{\alpha \beta}+2(k_c)^{\alpha \beta}-\frac{3}{2}\left(k^{WW}_F+k^{BB}_F\right)^{\alpha \beta}+\frac{1}{2}(k_{\phi \phi})^{\alpha \beta}+25(k_W)^{\alpha \beta}\right]\, ,
\end{equation}
\begin{equation}
\label{Z2}
(Z_{\hat{k}_F})_{\mu \alpha \beta \nu}^{\, \, \, \, \, \, \, \, \, \, \, \, \, \, \gamma \delta \eta \xi}(\hat{k}_F)_{ \gamma \delta \eta \xi}=
(\hat{k}_F)_{\mu \alpha \beta \nu}+N\frac{\alpha}{3\pi\epsilon}\left[(\hat{k}_F)_{\mu \alpha \beta \nu}-\frac{3}{4}\left(\hat{k}^{WW}_F+\hat{k}^{BB}_F\right)_{\mu \alpha \beta \nu}+9(\hat{k}_W)_{\mu \alpha \beta \nu} \right]\, ,
\end{equation}
where $(k_c)_{\alpha \beta}=\frac{1}{2}(c_{\alpha \beta}+c_{\beta \alpha})$.

Having calculated the $Z$ factors in the dimensional-regularization approach with minimal subtraction, we now turn to determine the corresponding beta functions. The beta functions for the tensor parameters $(k_F)_{\alpha \beta}$ and $(\hat{k}_F)_{\mu \alpha \beta \nu}$ are, by definition,
\begin{eqnarray}
(\beta_{k_F})_{\alpha \beta}&=&\mu \frac{d(k_F)_{\alpha \beta}}{d\mu} \, , \\
(\beta_{\hat{k}_F})_{\mu \alpha \beta \nu}&=&\mu \frac{d(\hat{k}_F)_{\mu \alpha \beta \nu}}{d\mu} \, ,
\end{eqnarray}
where $\mu$ is the mass unit of dimensional regularization. In the $\overline{MS}$-scheme the beta functions can be calculated directly from the simple $\epsilon$ pole in the corresponding $Z$ factor (see~\cite{QEDE} and references therein for details). In our case, we have
\begin{eqnarray}
\label{B1}
(\beta_{k_F})_{\alpha \beta}&=&2\alpha \frac{d(a^{k_F}_1)_{\alpha \beta}}{d\alpha} \, , \\
\label{B2}
(\beta_{\hat{k}_F})_{\mu \alpha \beta \nu}&=&2\alpha \frac{d(a^{\hat{k}_F}_1)_{\mu \alpha \beta \nu}}{d\alpha} \, ,
\end{eqnarray}
where $\alpha$ is the fine-structure constant, while $(a^{k_F}_1)_{\alpha \beta}$ and $(a^{\hat{k}_F}_1)_{\mu \alpha \beta \nu}$ are the coefficients of $1/\epsilon$ in Eqs.~(\ref{Z1}) and (\ref{Z2}), respectively. So, Eqs.~(\ref{B1}) and (\ref{B2}) lead to
\begin{eqnarray}
(\beta_{k_F})_{\alpha \beta}&=&N\frac{2\alpha}{3\pi \epsilon}\left[(k_F)_{\alpha \beta}+2(k_c)_{\alpha \beta}-\frac{3}{2}\left(k^{WW}_F+k^{BB}_F\right)_{\alpha \beta}+\frac{1}{2}(k_{\phi \phi})_{\alpha \beta}+25(k_W)_{\alpha \beta}\right]\, ,\\
(\beta_{\hat{k}_F})_{\mu \alpha \beta \nu}&=&N\frac{2\alpha}{3\pi\epsilon}\left[(\hat{k}_F)_{\mu \alpha \beta \nu}-\frac{3}{4}\left(\hat{k}^{WW}_F+\hat{k}^{BB}_F\right)_{\mu \alpha \beta \nu}+9(\hat{k}_W)_{\mu \alpha \beta \nu} \right]\, .
\end{eqnarray}

Given the beta functions, the next step is to study the behavior of the corresponding parameters in a wide range of energies by solving the renormalization-group equations. However, such study requires a calculation of all the beta functions associated with the mSME, which is lies beyond the scope of this work.

\section{Summary}
\label{C} The mSME provides us with a powerful tool to investigate effects from both {\it CPT} and LV in a model-independent fashion. Although the model contains all the interactions that are renormalizable, in the Dyson's sense, and which are compatible with observer Lorentz transformations and gauge symmetry, it involves subtleties that could lead to technical difficulties at the level of quantum fluctuations. Therefore, it is very  important to study the one-loop structure of the model, as many of their most interesting predictions can be derived at this level. This is, in part, the spirit of  this paper. We have presented an exact calculation of the photon propagator at one loop, in the context of the mSME. All the contributions from the extended Yukawa, Higgs, and Yang-Mills sectors were considered. This calculation, together with the one already given in the literature in the context of the QEDE, completes the one loop radiative correction to the photon propagator in the context of the mSME.

The issue of a gauge-fixing procedure that incorporates LV and its implications on the ghost sector was discussed. We based our discussion in the field-antifield formalism, in which the BRST symmetry arises naturally since the classical level, and allows us to quantize a wider variety of gauge systems. The quantization of the mSME does not really require of any additional ingredient to those used to quantize the SM, since both theories are governed by the same gauge group. However, the introduction of LV in a gauge-fixing procedure can considerably simplify loop calculations. With this in mind, we introduced nonlinear gauge-fixing functions for the electroweak group, as general as allowed by the BRST symmetry and renormalization theory, with ultimate goal of simplifying to the fullest the Higgs sector extension. Two nonlinear gauge-fixing procedures were discussed, one that does not  incorporate LV and another that does. In the latter case, we found that of the set of three tensors involved in the CPT-even Higgs sector extension, namely, $\{ (k_{\phi \phi})_{\alpha \beta}, (k_{\phi W})_{\alpha \beta}, (k_{\phi B})_{\alpha \beta}\}$, only the introduction of the symmetric one $(k_{\phi \phi})_{\alpha \beta}$ in the gauge-fixing functions leads to important simplifications in this sector. As an application, we calculated the contribution of $(k_{\phi \phi})_{\alpha \beta}$ to the photon propagator using both type of gauges, and showed, using dimensional regularization, that the divergent part of the corresponding amplitudes coincides. Since in dimensional regularization with minimal subtraction, the beta function for a given parameter can be calculated directly from the simple pole, this result simply reflects the fact that the beta functions are gauge-independent quantities. Explicit expressions for the ghost and gauge-fixing Lagrangians were presented, from which Feynman rules can be derived.

The most general flavor-violating structure of the Yukawa sector in both the lepton and quark sectors was considered. The most general Lagrangian for the pure-photon sector is characterized by the CS-type {\it CPT}-odd $(k_{AF})_\alpha$ vector and by the {\it CPT}-even Riemann-type tensor $(k_F)_{\mu \alpha \beta \nu}$. Since both the Yukawa sector and Higgs sector extensions are {\it CPT}-even, only the Riemann-type tensor $(k_F)_{\mu \alpha \beta \nu}$ receives one-loop contributions from these sectors. We have presented our discussion in terms of the ${\rm SO}(1,3)$ irreducible parts of $(k_F)_{\mu \alpha \beta \nu}$, namely, the Weyl-type tensor  $(\hat{k}_F)_{\mu \alpha \beta \nu}$, which is sensitive to birefringence; the Ricci-type symmetric tensor $(k_F)_{\alpha \beta}$, which defines the non-birefringent tensor $(\tilde{k}_F)_{\mu \alpha \beta \nu}$; and the curvature-type scalar $k_F$.

In the Yukawa sector, we found that all the irreducible parts of the Riemann-type tensor receive contributions. However, all these contributions are free of ultraviolet divergences, so such effects are unobservable, as they can be absorbed through a finite renormalization of the electromagnetic field, the electric charge,  $(\hat{k^f}_F)_{\mu \alpha \beta \nu}$, and $(k^f_F)_{\alpha \beta}$. The only observable  effect may come from the Ricci-type tensor $(k^f_F)_{\alpha \beta}$ and the scalar $k^f_F$, through a dimension-6 observer and gauge-invariant term. Such a dimension-6 term is finite, in the ultraviolet sense, in accordance with renormalization theory.

As far as the Higgs-sector extension is concerned, the contributions to the photon propagator are given by the two antisymmetric tensors $(k_{\phi W})_{\alpha \beta}, (k_{\phi B})_{\alpha \beta}$, and by the symmetric tensor $(k_{\phi \phi})_{\alpha \beta}$ as well. Due to gauge invariance, the antisymmetric contributions first arise at second order in these type of tensors, whereas symmetric tensors turn out to be linear in $(k_{\phi \phi})_{\alpha \beta}$. We performed exact calculations and discussed separately both types of contributions. In the case of the  antisymmetric contribution, we found that there are contributions to all irreducible parts of the corresponding Riemann-type tensors that can be constructed with products of tensors $ (k_{\phi W})_{\alpha \beta}, (k_{\phi B})_{\alpha \beta}$. These contributions, in contrast with those from the Yukawa sector, are divergent, so they have observable effects. Then, the sensitivity to birefringence of the Weyl-type tensor can be used to bound the tensors $ (k_{\phi W})_{\alpha \beta}, (k_{\phi B})_{\alpha \beta}$. To do this, we have used the reasonable assumption, by Colladay and Kosteleck\'y, that the one-loop effect from these tensors constitutes a radiative correction to the corresponding renormalized tensors and that it is of the same order of magnitude. We found that the bounds derived under this assumption are essentially the ones first obtained by Anderson-Sher-Turan~\cite{Sher} using a cut-off to extract the physical content from the divergent integrals. Also, we have found a finite contribution to the same dimension-6 observer and gauge-invariant term found in the Yukawa sector, which is in accordance with renormalization theory.

Regarding the impact of the symmetric $(k_{\phi \phi})_{\alpha \beta}$ tensor, it does not contribute to birefringence. We organized the corresponding contributions in terms of the scalar $k_{\phi \phi}$, the tensor $(k_{\phi \phi})_{\alpha \beta}$, and the non-birefringent tensor $(\tilde{k}_{\phi \phi})_{\mu \alpha \beta \nu}$. The contribution  proportional to $(k_{\phi \phi})_{\alpha \beta}$ corresponds to the one of the dimension-6 term already mentioned, which is free of ultraviolet divergences, in agreement with renormalization theory. On the other hand, both contributions proportional to $k_{\phi \phi}$ and $(\tilde{k}_{\phi \phi})_{\mu \alpha \beta \nu}$ are divergent. Consequently, there are observable effects via the non-birefringent $(\tilde{k}_{\phi \phi})_{\mu \alpha \beta \nu}$ part or, equivalently, the Ricci-type tensor $(k_{\phi \phi})_{\alpha \beta}$. We used the best available bounds on the components of the renormalized tensor $(k_{\phi \phi})_{\alpha \beta}$, obtained from the Laser Interferometer Gravitational-Wave Observatory, to bound  $(\tilde{k}_{\phi \phi})_{\mu \alpha \beta \nu}$ following the Colladay-Kosteleck\'y scheme. We found that the respective bounds are much stronger than those previously derived from properties of $\beta$ decay.

The contribution to the photon propagator of the Yang-Mills sector extension was also calculated. This contribution is characterized by the Riemann-type tensor $(k_W)_{\mu \alpha \beta \nu}$. We found that there are both birefringent and non-birefringent contributions characterized by the Weyl-type $(\hat{k}_W)_{\mu \alpha \beta \nu}$ tensor and the Ricci-type $(k_W)_{\alpha \beta}$ tensor, respectively. A finite contribution proportional to $(k_W)_{\alpha \beta}$ to the dimension-6 term already commented is also generated. These contributions are divergent, so there are observable effects. Bounds on the components of these tensors were discussed in the same spirit as the parameters of the Higgs sector.

The renormalization of the CPT-even photon propagator was discussed in dimensional regularization with minimal subtraction. We found that it is more natural to consider the Weyl-type $(\hat{k}_F)_{\mu \alpha \beta \nu}$ tensor, and the Ricci-type $(k_F)_{\alpha \beta}$ tensor, rather the Riemann-type $(k_F)_{\mu \alpha \beta \nu}$ tensor, as the fundamental parameters, since radiative corrections distinguish between one and the other. The corresponding beta functions $(\beta_{\hat{k}_F})_{\mu \alpha \beta \nu}$ and $(\beta_{k_F})_{ \alpha \beta}$ were derived.

\acknowledgments{We acknowledge financial support from CONACYT (M\' exico). J.M., H.N.S., and J.J.T. also acknowledge financial support from SNI (M\' exico). J.M. thanks support from C\' atedras CONACYT Program.}

\appendix

\section{The Field-Antifield formalism in the Standard Model}
\label{FAFF}At the classical level, BRST symmetry, which contains the essence  of the gauge symmetry, arises naturally from the {\it field-antifield formalism}~\cite{BRST2,BRST3}. The motivation to develop this formalism is to quantize gauge systems in a covariant way. It resorts to the construction of a configuration space endowed with a symplectic structure, called the {\it antibracket}, which resembles the graded Poisson parenthesis. This symplectic structure offers us, among other advantages, the powerful tool of canonical transformations. Our main goal is to construct, in this context, the most general ghost sector for the mSME, consistent with renormalization theory. Our starting point is the classical action for the mSME given in~\cite{SME}. The scalar and gauge sectors are strongly linked to each other in any theory involving spontaneous symmetry breaking, so we focus our discussion in such sectors. In this formalism, the ghost fields are recognized as genuine degrees of freedom that must enter the theory since the classical level. The original configuration space is extended to include a ghost field per each gauge parameter. It is assumed that ghost-fields statistics is opposite to the statistics of the corresponding gauge parameters, so, in the Yang-Mills case, they are fermion (anticommuting) fields. Let $\Phi^A$ generically denote the fields $G^{a\mu}, W^{i\mu}, B^\mu, {\cal C}^a,  {\cal C}^i,  {\cal C}, \phi$, which are, respectively, the gauge and ghost fields associated with the SM  $SU_C(3)\times SU_L(2)\times U_Y(1)$ gauge group, and the Higgs doublet. An additive conserved charge, called {\it ghost number}, is assigned to each field. Classical fields appearing in the classical action for the mSME have ghost number 0, whereas the ghost number of ghost fields is 1. This configuration space is further extended by introducing an antifield $\Phi^*_A$ per each field $\Phi^A$, that is, we have antifields $\Phi^*_A=G^*_{a\mu}, W^*_{i\mu}, B^*_\mu, {\cal C}^*_a,  {\cal C}^*_i,  {\cal C}^*, \phi^*$. Any antifield has opposite statistics to that of its associated field, while ghost number is given by $gh[\Phi^*_A]=-gh[\Phi^A]-1$. Even though antifields do not represent physical degrees of freedom, but are only a mathematical tool of the formalism, their introduction provides us with an unambiguous criterion for introducing a  gauge-fixing procedure, which is necessary to quantize the theory. It is worth commenting that antifields are not quantized; they are eliminated from the theory in a nontrivial way through a gauge-fixing procedure.

In the space of fields and antifields, the antibracket is defined through left and right derivatives as
\begin{equation}
(X,Y)=\frac{\delta_R X}{\delta \Phi^A}\frac{\delta_L Y}{\delta \Phi^*_A}-\frac{\delta_R X}{\delta \Phi^*_A}\frac{\delta_L Y}{\delta \Phi^A}\, ,
\end{equation}
where $X$ and $Y$ are arbitrary functionals on fields and antifields. Here the symbols $\delta_{L,R}/\delta\Phi^A$ and $\delta_{L,R}/\delta\Phi^*_A$ have been used to denote functional derivatives. The antibracket has properties~\cite{BRST3} that are similar to those of the graded version of the Poisson bracket~\cite{HT}. From this definition, we have, in particular, the {\it fundamental antibracket}
\begin{equation}
(\Phi^A,\Phi^*_B)=\delta^A_B \, ,
\end{equation}
so the fields and the antifields form canonical pairs, just as in  the usual phase space. An advantage of the field-antifield space is that all the symmetries of the  theory are manifest. In this formulation, the classical action is extended to include, besides ghosts, the antifields. This extended action, which is denoted by $S[\Phi,\Phi^*]$, is a bosonic functional with ghost number 0, which satisfies the {\it master equation}, given by
\begin{equation}
(S,S)=0\, .
\label{mastereq}
\end{equation}
The solution of the master equation plays a double role: on the one hand, it is the generating functional of the gauge structures of the system; and, on the other hand, it is the starting point to covariantly quantize the theory. Among the whole set of solutions of the master equation, only those which satisfy certain boundary conditions are interesting~\cite{BRST3}. One essential requirement is the classical limit $S[\Phi,\Phi^*]|_{\Phi^*=0}=S^{\rm SME}_0$; a solution that fulfills this is called a {\it proper solution}. Any proper solution can be expressed as a power series in the antifields (see~\cite{BRST3} for details).

It si well known that gauge symmetry cannot be realized in the quantum space. It is the BRST symmetry the one to be understood as a quantum symmetry~\cite{BRST1}. The presence of BRST symmetry in the field-antifield formalism, even before the implementation of some gauge-fixing procedure, is an important fact. The generator of the classical BRST symmetry is the proper solution $S$, through the antibracket. The classical BRST  transformation is given by
\begin{equation}
\delta_B X=(X,S)\, ,
\end{equation}
where $X$ is an arbitrary functional on fields and antifields. Thus the master equation, Eq.~(\ref{mastereq}), is nothing but the statement that the extended action $S$ is BRST invariant at the classical level. The BRST transformations of the field and antifields are:
\begin{eqnarray}
\label{BRSTF}
\delta_B \Phi^A&=&(\Phi^A,S)=\frac{\delta_L S}{\delta \Phi^*_A} \, , \\
\label{BRSTA}
\delta_B \Phi^*_A&=&(\Phi^*_A,S)=-\frac{\delta_L S}{\delta \Phi^A} \, .
\end{eqnarray}

The proper solution for pure Yang-Mills theories is well known~\cite{BRST3}. Here, we extend this solution to include spontaneous symmetry breaking. In the context of the mSME, the proper solution can be written as follows:
 \begin{eqnarray}
 \label{PSM}
 S[\Phi,\Phi^*]&=&\int d^4x \Big\{{\cal L}^{\rm SME}_{B}+G^*_{a\mu}{\cal D}^{ab\mu}{\cal C}^b+ \frac{1}{2}f^{abc}{\cal C}^*_a{\cal C}^b{\cal C}^c+
 +W^*_{i\mu}{\cal D}^{ij\mu}{\cal C}^j+ \frac{1}{2}\epsilon^{ijk}{\cal C}^*_i{\cal C}^j{\cal C}^k\nonumber \\
 &&+B^*_\mu \partial^\mu {\cal C}+ ig\left[\phi^{*\dag}\left(\frac{\sigma^i}{2}\right)\phi- \phi^\dag \left(\frac{\sigma^i}{2}\right)\phi^*\right]{\cal C}^i +ig'\left[\phi^{*\dag}\left(\frac{Y}{2}\right)\phi- \phi^\dag \left(\frac{Y}{2}\right)\phi^*\right]{\cal C}
 \Big\}\, ,
 \end{eqnarray}
 where $\phi^{*\dag}$ and $\phi^*$ are, respectively, the antidoublets (fermionic, with ghost number -1) of the Higgs doublets $\phi$ and $\phi^\dag$. Note that this action is bosonic and has ghost number 0. Also, $S$ must be real in the Grassmann's sense\footnote{This requirement is necessary to have a unitary $S$-matrix.}. Since all the gauge and ghost fields are real, this requirement is automatically satisfied in this part of $S$ if all the corresponding antifields are purely imaginary in the Grassmann's sense\footnote{Complex conjugation in the Grassmann's sense is defined as $(XY)^{\bigstar}=Y^\bigstar X^\bigstar$~\cite{HT}. For instance, $(G^*_{a\mu}{\cal C}^a)^\bigstar=({\cal C}^a)^\bigstar (G^*_{a\mu})^\bigstar={\cal C}^a (G^*_{a\mu})^\bigstar=G^*_{a\mu}{\cal C}^a$ only if the fermionic antifield $G^*_{a\mu}$ is purely imaginary.}. In the case of the Higgs doublet, this holds if the real and imaginary parts of the Higgs doublets $\phi$ and $\phi^\dag$ become purely imaginary in the Grassmann's sense in the corresponding antifields $\phi^*$ and $\phi^{*\dag}$. The Lagrangian ${\cal L}^{\rm SME}_{B}$, which is part of Eq.~(\ref{PSM}), contains, besides the Yang-Mills and Higgs sectors of the SM, the Lorentz-violating terms for these sectors, which are given in Ref.~\cite{SME}. In addition, ${\cal D}^{ab}_{\mu}=\delta^{ab}\partial_\mu-gf^{abc}G^c_\mu$ and ${\cal D}^{ij}_{\mu}=\delta^{ij}\partial_\mu-g\epsilon^{ijk}W^k_\mu$ are the covariant derivatives in the adjoint representations of the gauge groups $SU_C(3)$ and $ SU_L(2)$, respectively. It is interesting to show the BRST transformations that arise from Eqs.(\ref{BRSTF},\ref{BRSTA}). As an example, consider the quantum-chromodynamics group. For the fields we have
 \begin{eqnarray}
 \delta_B G^{a\mu}&=&{\cal D}^{ab\mu}{\cal C}^b\, , \\
 \delta_B{\cal C}^a&=&\frac{1}{2}f^{abc}{\cal C}^b{\cal C}^c\, ,
 \end{eqnarray}
while the corresponding antifield transformations are
\begin{eqnarray}
\delta_B G^*_{a\mu}&=&{\cal D}^{ab\nu}G^b_{\mu \nu}f^{abc}-f^{abc}G^*_{b\mu}{\cal C}^b \, ,\\
\delta_B{\cal C}^*_a&=&-{\cal D}^{ab\mu}G^*_{b\mu}+f^{abc}C^{*b}C^c\, .
\end{eqnarray}
Notice that the BRST transformations for the fields are independent of antifields, so they prevail even if the antifields are removed from the theory.

The extended action $S$ cannot be quantized directly because it is still gauge invariant. To define the corresponding path integral, one needs to introduce a gauge-fixing procedure to remove the degeneration associated to gauge symmetry. In addition, the path integral must be defined only in terms of fields, so the antifields must be removed from the theory before quantization. Notice that the antifields cannot simply be equated to 0, since this would lead us to the original theory, which is gauge invariant. Nevertheless, one can remove the antifields in a nontrivial way and simultaneously lift the degeneration from the theory. In other words, the choice of a mechanism to remove the antifields from the theory in a nontrivial manner is equivalent to pick a gauge-fixing procedure. Such a choice is, however, not arbitrary. Following Batalin and Vilkovsky~\cite{BV}, we introduce a fermionic functional $\Psi$ with ghost number $-1$. Then, the antifields are eliminated by defining
\begin{equation}
\label{GF}
\Phi^*_A=\frac{\delta \Psi[\Phi]}{\delta \Phi^A}\, ,
\end{equation}
where $\Psi $ depends on fields only. At this point it is not necessary to distinguish between left and right derivatives, as $\Psi$ is odd. The corresponding gauge-fixed action is $S_\Psi=S[\Phi^A,\delta \Psi/\delta \Phi^A]$. According to Eq.(\ref{GF}), $\Psi$ has ghost number -1, which cannot happen since all the fields of the theory have non-negative ghost number. To construct a $\Psi$ with the required ghost number, one introduces a {\it trivial pair} of fields $\{ B, \bar{\cal C}\}$~\cite{BRST3} per each gauge field, with ghost number $0$ and $-1$ respectively. By definition, the statistics of $B$ coincides with the statistics of the corresponding gauge parameter, whereas the statistics of $\bar{\cal C}$ is defined to be opposite to the statistics of such a gauge parameter. In Yang-Mills theories, $B$ and $\bar{\cal C}$ are even and odd fields, respectively. The bosonic field $B$ is usually called {\it auxiliary field}, whereas the $\bar{\cal C}$ field is the Faddeev-Popov antighost of ${\cal C}$. The proper solution to the master equation is unique up to a canonical transformation and the addition of trivial pairs:
\begin{equation}
\label{T}
S_{\rm trivial}=\int d^4x \left(B^a \bar{\cal C}^*_a+B^i \bar{\cal C}^*_i+B \bar{\cal C}^*\right)\, ,
\end{equation}
where $\bar{\cal C}^*$ is the antifield of $\bar{\cal C}$. Since the auxiliary field $B$ is assumed to be real, the antifield $\bar{\cal C}^*$ must also be real, so the trivial action $S_{\rm trivial}$ is a real quantity, which is correct. The proper solution of the master equation is obtained by summing Eqs.~(\ref{PSM}) and (\ref{T}) together:
\begin{eqnarray}
\label{PS}
 S&=&\int d^4x \Big\{{\cal L}^{\rm SME}_{B}+G^*_{a\mu}{\cal D}^{ab\mu}{\cal C}^b+ \frac{1}{2}f^{abc}{\cal C}^*_a{\cal C}^b{\cal C}^c+
 +W^*_{i\mu}{\cal D}^{ij\mu}{\cal C}^j+ \frac{1}{2}\epsilon^{ijk}{\cal C}^*_i{\cal C}^j{\cal C}^k+B^*_\mu \partial^\mu {\cal C}\nonumber \\
 &&+ig\left[\phi^{*\dag}\left(\frac{\sigma^i}{2}\right)\phi- \phi^\dag \left(\frac{\sigma^i}{2}\right)\phi^*\right]{\cal C}^i+ig'\left[\phi^{*\dag}\left(\frac{Y}{2}\right)\phi- \phi^\dag \left(\frac{Y}{2}\right)\phi^*\right]{\cal C}+B^a \bar{\cal C}^*_a+B^i \bar{\cal C}^*_i+B \bar{\cal C}^*\Big\}\, .
\end{eqnarray}
Some comments concerning the SME matter fields are in order here. Antifields for all leptons and quarks can be introduced in the same way as it was done for gauge, scalar, and ghost fields. The BRST transformation given in Eq.~(\ref{BRSTF}) simply reproduces the form in which these fields transform under the SM group, with gauge parameters replaced by ghost fields. In contrast with the boson sector, all the antifields of leptons and quarks can be removed from the theory, in a trivial way, by simply putting them equal to zero. It is worth keeping in mind that the same cannot be done with scalar fields: the Higgs mechanism links the scalar fields to the gauge sector, so the Higgs sector affects the ghost sector, which must be taken into account to fix the gauge and carry out quantization.

We now proceed to remove the antifields from the proper solution, Eq.~(\ref{PS}), by introducing an appropriate gauge-fermion $\Psi$. Since this functional defines the structure of the ghost sector, we will proceed as general as possible. We choose a $\Psi$ of the way
\begin{equation}
\Psi=\int d^4x\left\{\bar{\cal C}^a\left(f^a+\frac{\hat{\xi}}{2}B^a+f^{abc}\bar{\cal C}^b C^c\right)+
\bar{\cal C}^i\left(f^i+\frac{\xi}{2}B^i+\epsilon^{ijk}\bar{\cal C}^j C^k\right)+\bar{\cal C}\left(f+\frac{\xi}{2}B\right)\right\}\, ,
\label{gaugeff}
\end{equation}
where $\hat{\xi}$ and $\xi$ are the gauge real parameters for the quantum-chromodynamics and electroweak groups, respectively. For the sake of simplicity, we have used the same gauge parameter for the groups $SU_L(2)$ and $U_Y(1)$. The $f^a$, $f^i$, and $f$ are, in general, real functions of gauge and scalar fields. Therefore, the functional $\Psi$, so defined, is fermionic and has ghost number $-1$. Since the antighost fields $\bar{\cal C}^a$, $\bar{\cal C}^i$, and $\bar{\cal C}$ are imaginary, this functional is purely imaginary in the Grassmann's sense. The {\it gauge-fixing functions} $f^a$, $f^i$, and $f$ are quite arbitrary, but note that they are restricted by the power-counting criterion of renormalization theory. To eliminate the antifields, we do not need to specify these functions, which makes sense, since many different physically equivalent gauge-fixing procedures, leading to the same $S$-matrix elements, are available. However, it is important to stress that, in general, off-shell Green's functions are gauge-fixing dependent. With the gauge-fermion $\Psi$ defined in Eq.~(\ref{gaugeff}), Eq.~(\ref{GF}) yields the following expressions for the antifields:
\begin{eqnarray}
G^*_{a\mu}&=&\bar{\cal C}^b\frac{\partial f^b}{\partial G^{a\mu}}\, , \ \ \ \ \  \bar{\cal C}^*_a=f^a+\frac{\hat{\xi}}{2}B^a+f^{abc}\bar{\cal C}^b C^c\, , \ \ \ \ {\cal C}^*_a=f^{abc}\bar{\cal C}^b\bar{\cal C}^c\, , \ \ \ \ B^*_a=\frac{\hat{\xi}}{2}\bar{\cal C}^a\, , \\
W^*_{i\mu}&=&\bar{\cal C}^j\frac{\partial f^j}{\partial W^{i\mu}}\, , \ \ \ \ \  \bar{\cal C}^*_i=f^i+\frac{\xi}{2}B^i+\epsilon^{ijk}\bar{\cal C}^j C^k\, , \ \ \ \ \ {\cal C}^*_i=\epsilon^{ijk}\bar{\cal C}^j\bar{\cal C}^k\, , \  \ \ \ \ B^*_i=\frac{\xi}{2}\bar{\cal C}^i\, , \\
B^*_{\mu}&=&\bar{\cal C}\frac{\partial f}{\partial B^{\mu}}\, , \ \ \ \ \ \ \ \bar{\cal C}^*=f+\frac{\xi}{2}B\, , \ \ \ \ \ \ \ \ \ \ \ \ \ \ \ \ \ \ \ \ \ \ \ \ {\cal C}^*=0\, , \ \ \ \ \ \ \ \ \ \ \ \ \  B^*=\frac{\xi}{2}\bar{\cal C}\, ,\\
\phi^{*\dag}_i&=&\bar{\cal C}^j \frac{\partial f^j}{\partial \phi_i}+\bar{\cal C}\frac{\partial f}{\partial \phi_i}\, , \ \ \ \ \phi^{*}_i=\bar{\cal C}^j \frac{\partial f^j}{\partial \phi^\dag_i}+\bar{\cal C}\frac{\partial f}{\partial \phi^\dag_i}\, .
\end{eqnarray}
Due to the fact that the sets $\{ G^{a\mu},W^{i\mu},B^{\mu}, {\cal C}^a,{\cal C}^i,{\cal C}, B^a,  B^i,  B, \bar{C}^*_a,  \bar{C}^*_i,  \bar{C}^*\}$ and $\{G^*_{a\mu}, W^*_{i\mu}, B^*_{\mu},{\cal C}^*_a,{\cal C}^*_i,{\cal C}^*, B^*_a, B^*_i,B^*,\bar{\cal C}^a,  \bar{\cal C}^i, \bar{\cal C} \}$ are, respectively, constituted by real and imaginary fields in the Grassmann's sense, the above mathematical relations require the gauge-fixing functions $f^a$, $f^i$, and $f$ to be real. We stress this fact because we are interested in constructing Lorentz-violating gauge-fixing functions within a framework as general as possible, so that the presence of background complex tensors is allowed.

Once eliminated the antifields from the BRST-invariant proper solution, Eq.~(\ref{PS}), we arrive to the gauge-fixed BRST action,
\begin{eqnarray}
\label{GFS}
S_{\Psi}&=&\int d^4x \Big\{{\cal L}_{\rm SME}+\bar{\cal C}^a\frac{\partial f^a}{\partial G^b_\mu}{\cal D}^{bc}_\mu {\cal C}^c+
\bar{\cal C}^i\frac{\partial f^i}{\partial W^j_\mu}{\cal D}^{jk}_\mu {\cal C}^k+\bar{\cal C}\frac{\partial f}{\partial B_\mu}\partial_\mu {\cal C} +\frac{1}{2}f^{abc}f^{ade}\,\bar{\cal C}^d \bar{\cal C}^e{\cal C}^b {\cal C}^c + \frac{1}{2}\epsilon^{ijk}\epsilon^{ilm}\,\bar{\cal C}^l \bar{\cal C}^m{\cal C}^j {\cal C}^k\nonumber \\
&&+ \frac{\hat{\xi}}{2}B^aB^a+B^a\left(f^a+f^{abc}\bar{\cal C}^b{\cal C}^c\right)+ \frac{\xi}{2}B^iB^i+B^i\left(f^i+\epsilon^{ijk}\bar{\cal C}^j{\cal C}^k\right)+\frac{1}{\xi}B^2+Bf+{\cal L}_{\rm GH}\Big\}\, ,
\end{eqnarray}
where the classical Lagrangian ${\cal L}_{\rm SME}$ already includes all the sectors of the mSME. In the above expression, ${\cal L}_{\rm GH}$ contains the ghost-Higgs interactions, and is given by
\begin{eqnarray}
\label{GHi}
{\cal L}_{\rm GH}&=&ig\bar{\cal C}^i\left[\frac{\partial f^i}{\partial \phi_k}\left(\frac{\sigma^j}{2}\right)_{kl}\phi_l-\phi^\dag_l\left(\frac{\sigma^j}{2}\right)_{lk} \frac{\partial f^i}{\partial \phi^\dag_k}  \right]{\cal C}^j+ ig\bar{\cal C}\left[\frac{\partial f}{\partial \phi_k}\left(\frac{\sigma^j}{2}\right)_{kl}\phi_l-\phi^\dag_l\left(\frac{\sigma^j}{2}\right)_{lk} \frac{\partial f}{\partial \phi^\dag_k}  \right]{\cal C}^j\nonumber \\
&&+ig'\bar{\cal C}^i\left[\frac{\partial f^i}{\partial \phi_j}\left(\frac{1}{2}\right)\phi_j-\phi^\dag_j\left(\frac{1}{2}\right) \frac{\partial f^i}{\partial \phi^\dag_j}  \right]{\cal C}+ig'\bar{\cal C}\left[\frac{\partial f}{\partial \phi_i}\left(\frac{1}{2}\right)\phi_i-\phi^\dag_i\left(\frac{1}{2}\right) \frac{\partial f}{\partial \phi^\dag_i}  \right]{\cal C}\, .
\end{eqnarray}
The action $S_{\Psi}$ is invariant under the gauge-fixed BRST transformations~\cite{BRST1}:
\begin{eqnarray}
\delta_{B_{\Psi}}G^a_\mu&=&{\cal D}^{ab}_\mu {\cal C}^b\, , \ \ \ \  \delta_{B_{\Psi}}{\cal C}^a=\frac{1}{2}f^{abc}{\cal C}^b{\cal C}^c\, , \ \ \ \ \delta_{B_{\Psi}}\bar{\cal C}^a=B^a \, ,  \ \ \ \ \delta_{B_{\Psi}}B^a=0\, ,\\
\delta_{B_{\Psi}}W^i_\mu&=&{\cal D}^{ij}_\mu {\cal C}^j\, , \ \ \ \  \delta_{B_{\Psi}}{\cal C}^i=\frac{1}{2}\epsilon^{ijk}{\cal C}^j{\cal C}^k\, , \ \ \ \ \delta_{B_{\Psi}}\bar{\cal C}^i=B^i \, ,  \ \ \ \ \delta_{B_{\Psi}}B^i=0\, ,\\
\delta_{B_{\Psi}}B_\mu&=&\partial_\mu {\cal C}\, , \ \ \ \ \ \ \ \  \delta_{B_{\Psi}}{\cal C}=0\, , \ \ \ \ \ \ \ \ \ \ \ \ \ \ \ \ \delta_{B_{\Psi}}\bar{\cal C}=B \, , \ \  \ \ \ \ \delta_{B_{\Psi}}B=0\, ,
\end{eqnarray}
\begin{equation}
\delta_{B_{\Psi}}\phi=ig\left(\frac{\sigma^i}{2}\right)\phi \, {\cal C}^i+ig'\left(\frac{Y}{2}\right)\phi \, {\cal C}\, .
\end{equation}
Notice that $\frac{\partial f^a}{\partial G^b_\mu}{\cal D}^{bc}_\mu {\cal C}^c=\delta_{B_\Psi}f^a$, etc. On the other hand, the auxiliary fields $B^a$, $B^i$, and $B$ appear quadratically in Eq.~(\ref{GFS}), so they can be integrated out from the path integral. Since the coefficients of the quadratic terms do not depend on the fields, their integration is equivalent to applying the equations of motion to the gauge-fixed BRST action. Once eliminated these fields, one obtains the effective action given by Eq.~(\ref{EQA}).

\section{Fermionic form factors}
\label{FF} The form factors of the fermionic contribution to the photon propagator are:
\begin{equation}
f_1(q^2)=\frac{16v^2m_Am_B}{(m^2_A-m^2_B)^2}\left[B_0(3)+B_0(4)-2B_0(5)\right]\, ,
\end{equation}
\begin{eqnarray}
f_2(q^2)&=&\frac{32iv^2}{3(m^2_A-m^2_B)^2(4m^2_A-q^2)(4m^2_A-q^2)}\Big\{\left[B_0(3)+B_0(4)-2B_0(5)\right]q^6 +2\Big[m^2_A\left(3B_0(5)-B_0(3)-2B_0(4)\right)\nonumber \\
&&+m^2_B\left(3B_0(5)-B_0(4)-2B_0(3)\right)\Big]q^4-2(m^2_A-m^2_B)\Big[m^2_A\left(2B_0(1)+4B_0(3)-6B_0(5)+5\right)\nonumber \\
&&-m^2_B\left(2B_0(2)+4B_0(4)-6B_0(5)+5\right)\Big]q^2+2\Big[ 2m^6_A\left(B_0(1)+3B_0(3)-4B_0(5)+4\right)\nonumber \\
&&-3m^5_Am_B\left(B_0(1)-B_0(3)+2\right)+2m^4_Am^2_B\left(3B_0(1)-4B_0(2)+5B_0(3)-4B_0(5)-4\right)\nonumber \\
&&+3m^3_Am^3_B\left(B_0(1)+B_0(2)-B_0(3)-B_0(4)+4\right) -2m^2_Am^4_B\left(4B_0(1)-3B_0(2)+4B_0(5)-5B_0(4)+4\right)\nonumber \\
&&-3m_Am^5_B\left(B_0(2)-B_0(4)+2\right)+2m^6_B\left(B_0(2)+3B_0(4)-4B_0(5)+4\right)\Big] \nonumber \\
&&-\frac{8m^2_Am^2_B(m^2_A-m^2_B)}{q^2}\Big[2m^2_A\left(B_0(1)+3B_0(3)-4B_0(5)+4\right)-3m_Am_B\left(B_0(1)-B_0(2)-B_0(3)+B_0(4)\right)\nonumber \\
&&-2m^2_B\left(B_0(2)+3B_0(4)-4B_0(5)\right) \Big]\Big\}\, ,
\end{eqnarray}

\begin{eqnarray}
f_3(q^2)&=&\frac{16iv^2}{3(m^2_A-m^2_B)^2(4m^2_A-q^2)(4m^2_A-q^2)}\Big\{\left[B_0(3)+B_0(4)-2B_0(5) \right]q^6\nonumber \\
&&+2\left[m^2_A\left(3B_0(5)-B_0(3)-2B_0(4)\right) +3m_Am_B\left(B_0(3)+B_0(4)-2B_0(5)\right)+m^2_B\left(3B_0(5)-2B_0(3)-B_0(4)\right)\right]q^4\nonumber \\
&&+2\Big[m^4_A\left(6B_0(5)-2B_0(1)-4B_0(3)-5\right) -12m^3_Am_B\left(B_0(3)+B_0(4)-2B_0(5) \right) +2m^2_Am^2_B\Big(B_0(1)+B_0(2)\nonumber \\
&&+2B_0(3)+2B_0(4)-6B_0(5)+5 \Big)-12m_Am^3_B\left(B_0(3)+B_0(4)-2B_0(5) \right)+m^4_B\Big(6B_0(5)-2B_0(2)-4B_0(4)\nonumber \\
&&-5 \Big)\Big]q^2 +4\Big[m^6_A\left(B_0(1)+3B_0(3)-4B_0(5)+4\right) -3m^5_Am_B\left(B_0(1)-B_0(3)+2 \right)+m^4_Am^2_B\Big(3B_0(1)-4B_0(2)\nonumber \\
&&+5B_0(3)-4B_0(5)-4 \Big) +3m^3_Am^3_B \left(B_0(1)+B_0(2)+7B_0(3)+7B_0(4)-16B_0(5)+4  \right)+m^2_Am^4_B\Big(-4B_0(1)\nonumber \\
&&+3B_0(2)+5B_0(4)-4B_0(5)-4 \Big)-3m_Am^5_B\left(B_0(2)-B_0(4)+2\right)+m^6_B\left(B_0(2)+3B_0(4)-4B_0(5)+4 \right)\Big]\nonumber \\
&&-\frac{16m^2_Am^2_B(m^2_A-m^2_B)}{q^2}\Big[m^2_A\left(B_0(1)+3B_0(3)-4B_0(5)+4 \right)-3m_Am_B\left(B_0(1)-B_0(2)-B_0(3)+B_0(4) \right)\nonumber \\
&&-m^2_B\left(B_0(1)+3B_0(4)-4B_0(5)+4 \right) \Big]  \Big\}\, ,
\end{eqnarray}

\begin{eqnarray}
f_4(q^2)&=&\frac{16v^2}{3(m^2_A-m^2_B)^2}\Big\{\left[B_0(3)+B_0(4)-2B_0(5) \right]q^2+2\Big[ m^2_A\left(B_0(4)-B_0(5) \right)\nonumber \\
&&-3m_Am_B\left(B_0(3)+B_0(4)-2B_0(5) \right)+m^2_B\left(B_0(3)-B_0(5) \right)\Big] \nonumber \\
&&-\frac{4(m^2_A-m^2_B)}{q^2}\left[ m^2_A\left(B_0(3)-B_0(5)+1 \right)-m^2_B\left(B_0(4)-B_0(5)+1 \right)\right] \Big\}\, ,
\end{eqnarray}

\begin{eqnarray}
f_5(q^2)&=&\frac{16v^2}{3(m^2_A-m^2_B)^2}\Big\{\left[2B_0(5)-B_0(3)-B_0(4)\right]q^2+2\Big[ m^2_A\left(B_0(5)-B_0(4) \right)\nonumber \\
&&-3m_Am_B\left(B_0(3)+B_0(4)-2B_0(5) \right)+m^2_B\left(B_0(5)-B_0(3) \right)\Big] \nonumber \\
&&+\frac{4(m^2_A-m^2_B)}{q^2}\left[ m^2_A\left(B_0(3)-B_0(5)+1 \right)-m^2_B\left(B_0(4)-B_0(5)+1 \right)\right] \Big\}\, ,
\end{eqnarray}

\begin{eqnarray}
f_6(q^2)&=&\frac{16iv^2}{3(m^2_A-m^2_B)^2(4m^2_A-q^2)(4m^2_A-q^2)}\Big\{\left[B_0(3)+B_0(4)-2B_0(5) \right]q^6\nonumber \\
&&-2\left[m^2_A\left(B_0(3)+2B_0(4)-3B_0(5) \right)+ m^2_B\left(B_0(4)+2B_0(3)-3B_0(5) \right)\right]q^4\nonumber \\
&&-2\Big[m^4_A\left(2B_0(1)+4B_0(3)-6B_0(5)+5 \right)+m^4_B\left(2B_0(2)+4B_0(4)-6B_0(5)+5 \right)\nonumber \\
&&-2m^2_Am^2_B\left(B_0(1)+B_0(2)+2B_0(3)+2B_0(4)-6B_0(5)+5\right) \Big]q^2\nonumber \\
&&4\Big[m^6_A\left(B_0(1)+3B_0(3)-4B_0(5)+4 \right)+m^4_Am^2_B\left(3B_0(1)-4B_0(2)+5B_0(3)-4B_0(5)-4 \right)\nonumber \\
&&+m^2_Am^4_B\left(-4B_0(1)+3B_0(2)+5B_0(4)-4B_0(5)-4 \right)+m^6_B\left(B_0(2)+3B_0(4)-4B_0(5)+4 \right) \Big] \nonumber \\
&&-\frac{16m^2_Am^2_B(m^2_A-m^2_B)}{q^2}\left[m^2_A\left(B_0(1)+3B_0(3)-4B_0(5)+4 \right) - m^2_B\left(B_0(2)+3B_0(4)-4B_0(5)+4 \right)\right]\Big\}\, ,
\end{eqnarray}

\begin{eqnarray}
f_7(q^2)&=&\frac{16iv^2}{3(m^2_A-m^2_B)^2(4m^2_A-q^2)(4m^2_A-q^2)}\Big\{\left[B_0(3)+B_0(4)-2B_0(5) \right]q^6-2\Big[m^2_A\left(B_0(3)+2B_0(4)-3B_0(5) \right)\nonumber \\
&&-3m_Am_B\left(B_0(3)+B_0(4)-2B_0(5) \right) +m^2_B\left(2B_0(3)+B_0(4)-3B_0(5) \right)\Big]q^4-2\Big[m^4_A\Big(2B_0(1)+4B_0(3)\nonumber \\
&&-6B_0(5)+5 \Big)+12m^3_Am_B\left(B_0(3)+B_0(4)-2B_0(5) \right) -2m^2_Am^2_B\Big(B_0(1)+B_0(2)+2B_0(3)+2B_0(4)\nonumber \\
&&-6B_0(5)+5 \Big)+12m_Am^3_B\left(B_0(3)+B_0(4)-2B_0(5) \right)+m^4_B\left(2B_0(2)+4B_0(4)-6B_0(5)+5 \right)\Big]q^2\nonumber \\
&&+4\Big[m^6_A\left(B_0(1)+3B_0(3)-4B_0(5)+4 \right)-3m^5_Am_B\left(B_0(1)-B_0(3)+2 \right) +m^4_Am^2_B\Big(3B_0(1)-4B_0(2)\nonumber \\
&&+5B_0(3)-4B_0(5)-4 \Big)+3m^3_Am^3_B\left(B_0(1)+B_0(2)+7B_0(3)+7B_0(4)-16B_0(5) +4\right)\nonumber \\
&&+m^2_Am^4_B\left(-4B_0(1)+3B_0(2)+5B_0(4)-4B_0(5)-4 \right)-3m_Am^5_B\left(B_0(2)-B_0(4)+2 \right)\nonumber \\
&&+m^6_B\left(B_0(2)+3B_0(4)-4B_0(5) +4\right)\Big]-\frac{16m^2_Am^2_B(m^2_A-m^2_B)}{q^2}\Big[m^2_A\left(B_0(1)+3B_0(3)-4B_0(5)+4 \right)\nonumber \\
&&-3m_Am_B\left(B_0(1)-B_0(2)-B_0(3)+B_0(4) \right)- m^2_B\left(B_0(2)+3B_0(4)-4B_0(5)+4 \right)\Big] \Big\}\, ,
\end{eqnarray}

\begin{eqnarray}
f_8(q^2)&=&\frac{16v^2}{3(m^2_A-m^2_B)^2(4m^2_A-q^2)(4m^2_A-q^2)}\Big\{\Big[m^2_A\left(B_0(4)-B_0(3) \right)+2m_Am_B\left(B_0(3)+B_0(4)-2B_0(5) \right)\nonumber \\
&&+m^2_B\left(B_0(3)-B_0(4) \right)\Big]q^4+\Big[m^4_A\left(2B_0(1)+2B_0(3)-4B_0(4)+3 \right)-8m^3_Am_B\left(B_0(3)+B_0(4)-2B_0(5) \right)\nonumber \\
&&-2m^2_Am^2_B\left(B_0(1)+B_0(2)-B_0(3)-B_0(4)+3 \right)-8m_Am^3_B\left(B_0(3)+B_0(4)-2B_0(5) \right) \nonumber \\
&&+m^4_B\left(2B_0(2)-4B_0(3)+2B_0(4) \right)\Big]q^2+2\Big[m^6_A\left(B_0(3)-B_0(1) \right)+3m^5_Am_B\left(B_0(1)-B_0(3)+2 \right) \nonumber \\
&&+m^4_Am^2_B\left(4B_0(2)-3B_0(1)+4B_0(4)-5B_0(3) \right)+m^3_Am^3_B\Big(-3B_0(1)-3B_0(2)+19B_0(3)+19B_0(4)\nonumber \\
&&-32B_0(5)-12\Big)+m^2_Am^4_B\left(4B_0(1)-3B_0(2)+4B_0(3)-5B_0(4) \right)+3m_Am^5_B\left(B_0(2)-B_0(4)+2 \right)\nonumber \\
&&+m^6_B\left(B_0(4)-B_0(2) \right)\Big]+\frac{8m^2_Am^2_B(m^2_A-m^2_B)}{q^2}\Big[m^2_A\left(B_0(1)-B_0(3)\right)+m^2_B\left(B_0(4)-B_0(2) \right)\nonumber \\
&&-3m_Am_B\left(B_0(1)-B_0(2)-B_0(3)+B_0(4)  \right) \Big] \Big\}\, ,
\end{eqnarray}
\begin{eqnarray}
f_9(q^2)&=&\frac{16v^2}{3(m^2_A-m^2_B)^2(4m^2_A-q^2)(4m^2_A-q^2)}\Big\{\Big[m^2_A\left(B_0(3)-B_0(4) \right) +m^2_B\left(B_0(4)-B_0(3) \right)\nonumber \\
&&2m_Am_B\left(B_0(3)+B_0(4)-2B_0(5) \right)\Big]q^4+\Big[m^4_A\left(4B_0(4)-2B_0(1)-2B_0(3)-3 \right)\nonumber \\
&&-8m^3_Am_B\left(B_0(3)+B_0(4)-2B_0(5) \right)+2m^2_Am^2_B\left(B_0(1)+B_0(2)-B_0(3)-B_0(4)+3 \right)\nonumber \\
&&-8m_Am^3_B\left(B_0(3)+B_0(4)-2B_0(5) \right)+m^4_B\left(4B_0(3)-2B_0(2)-2B_0(4)-3 \right) \Big]q^2\nonumber \\
&&+2\Big[m^6_A\left(B_0(1)-B_0(3)\right)+3m^5_Am_B\left(B_0(1)-B_0(3)+2 \right) +m^4_Am^2_B\left(3B_0(1)-4B_0(2)+5B_0(3)-4B_0(4) \right)\nonumber\\
&&+m^3_Am^3_B\left(-3B_0(1)-3B_0(2)+19B_0(3)+19B_0(4)-32B_0(5)-12 \right) +m^2_Am^4_B\Big(-4B_0(1)+3B_0(2)\nonumber \\
&&-4B_0(3)+5B_0(4) \Big)+3m_Am^5_B\left(B_0(2)-B_0(4)+2 \right)+m^6_B\left(B_0(2)-B_0(4) \right)\Big]\nonumber \\
&&-\frac{8m^2_Am^2_B(m^2_A-m^2_B)}{q^2}\Big[m^2_A\left(B_0(1)-B_0(3)\right)+3m_Am_B\left(B_0(1)-B_0(2)-B_0(3)+B_0(4)  \right) \nonumber \\
&&+m^2_B\left(B_0(4)-B_0(2) \right)\Big]  \Big\}\, ,
\end{eqnarray}
\begin{eqnarray}
f_{10}(q^2)&=&\frac{16iv^2}{9(m^2_A-m^2_B)^2(4m^2_A-q^2)(4m^2_A-q^2)}\Big\{\left[2B_0(5)-B_0(3)-B_0(4) \right]q^6 +2\Big[m^2_A(B_0(3)+2B_0(4)-3B_0(5))\nonumber \\
&&+m^2_B(2B_0(3)+B_0(4)-3B_0(5)) \Big]q^4+2\Big[m^4_A(2B_0(1)+4B_0(3)-6B_0(5)+5)-2m^2_Am^2_B\Big(B_0(1)+B_0(2)\nonumber \\
&&+2B_0(3)+2B_0(4)-6B_0(5)+5 \Big)+m^4_B(2B_0(2)+4B_0(4)-6B_0(5)+5) \Big]q^2-2\Big[2m^6_A\Big(B_0(1)+3B_0(3)\nonumber \\
&&-4B_0(5)+4 \Big) -3m^5_Am_B(B_0(1)-B_0(3)+2)+2m^4_Am^2_B(3B_0(1)-4B_0(2)+5B_0(3)-4B_0(5)-4)\nonumber \\
&&+3m^3_Am^3_B(B_0(1)+B_0(2)-B_0(3)-B_0(4)+4) -2m^2_Am^4_B(4B_0(1)-3B_0(2)-5B_0(4)+4B_0(5)+4)\nonumber \\
&&-3m_Am^5_B(B_0(2)-B_0(4)+2)+2m^6_B(B_0(2)+3B_0(4)-4B_0(5)+4)\Big]\nonumber \\
&&+\frac{8m^2_Am^2_B(m^2_A-m^2_B)}{q^2}\Big[2m^2_A(B_0(1)+3B_0(3)-4B_0(5)+4) -2m^2_B(B_0(2)+3B_0(4)-4B_0(5)+4)\nonumber \\
&&-3m_Am_B(B_0(1)-B_0(2)-B_0(3)+B_0(4)) \Big]\Big\}\, .
\end{eqnarray}
The various $B_0(i)$ terms appearing in the above form factors is a shorthand notation for two-point Passarino-Veltman scalar functions~\cite{PV}, given by:
\begin{eqnarray}
B_0(1)&=&B_0(0,m^2_A,m^2_A)\, , \\
B_0(2)&=&B_0(0,m^2_B,m^2_B)\, , \\
B_0(3)&=&B_0(q^2,m^2_A,m^2_A)\, , \\
B_0(4)&=&B_0(q^2,m^2_B,m^2_B)\, , \\
B_0(5)&=&B_0(q^2,m^2_A,m^2_B)\, ,
\end{eqnarray}
where we have used the definition of these functions given in Ref.~\cite{FeynCalc}, which, in $D$ dimensions, is:
\begin{equation}
B_0(q^2,m^2_1,m^2_1)=\frac{1}{i\pi^2}\int d^Dk\frac{1}{[k^2-m^2_1][(k+q)^2-m^2_2]}\, .
\end{equation}
The general solution of this function is well known in the literature~\cite{Denner}, however, for clarity purposes, we reproduce it here:
\begin{equation}
\label{GBS}
B_0(q^2,m^2_1,m^2_1)=\frac{1}{2}\Delta_{m_1}+\frac{1}{2}\Delta_{m_2}+1-\frac{m^2_1+m^2_2}{m^2_1-m^2_2}\log\left(\frac{m_1}{m_2}\right)+F(q^2,m^2_1,m^2_2)\, ,
\end{equation}
where $\Delta_{m_i}=\frac{1}{\epsilon}-\gamma_E+\log(4\pi)-\log\left(\frac{m^2_i}{\mu^2}\right)$, with $\epsilon=2-\frac{D}{2}$, $\gamma_E$ the  Euler-Mascheroni constant, and $\mu$ the mass scale of the dimensional regularization scheme. In addition,

\begin{equation}
F(q^2,m^2_1,m^2_2)= 1+\left(\frac{m^2_1-m^2_2}{q^2}-\frac{m^2_1+m^2_2}{m^2_1-m^2_2}\right)\log\left(\frac{m_2}{m_1}\right) +\hat{F}(q^2,m^2_1,m^2_2) \, ,
\end{equation}
with

\begin{align}
&\hat{F}(q^2,m^2_1,m^2_2)=
&\left \{ \begin{array}{l}
\\ \displaystyle{\frac{\sqrt{(m_1+m_2)^2-q^2}\sqrt{(m_1-m_2)^2-q^2}}{q^2}\log\left(\frac{\sqrt{(m_1+m_2)^2-q^2}+\sqrt{(m_1-m_2)^2-q^2}}
{\sqrt{(m_1+m_2)^2-q^2}-\sqrt{(m_1-m_2)^2-q^2}}\right)}
\\
\\
\mbox{for } q^2<(m_1-m_2)^2
\\
\\
\\
\displaystyle{-2\frac{\sqrt{(m_1+m_2)^2-q^2}\sqrt{q^2-(m_1-m_2)^2}}{q^2}\arctan\left(\frac{\sqrt{q^2-(m_1-m_2)^2}}{\sqrt{(m_1+m_2)^2-q^2}}\right)}
\\
\\
\mbox{for } (m_1-m_2)^2<q^2<(m_1+m_2)^2
\\
\\
\\
\displaystyle{-\frac{\sqrt{q^2-(m_1+m_2)^2}\sqrt{q^2-(m_1-m_2)^2}}{q^2}\left[\log\left(\frac{\sqrt{q^2-(m_1+m_2)^2}+\sqrt{q^2-(m_1-m_2)^2}}
{\sqrt{q^2-(m_1-m_2)^2}-\sqrt{q^2-(m_1+m_2)^2}}\right)-i\pi\right]}
\\
\\
\mbox{for } q^2>(m_1+m_2)^2 \, .
\\
\\
\end{array}\right.
\end{align}
The $F(q^2,m^2_1,m^2_2)$ function is defined in such a way that it disappears in the limit when $q^2$ tends to zero. As it can be appreciated from the above expressions, the form factors $f_i(q^2)$ depend, in addition, on $B_0(i)$ functions of the way $B_0(q^2,m^2,m^2)$ and $B_0(0,m^2,m^2)$. In the former case, the general solution (\ref{GBS}) acquires a simpler form, but it is really simple in  the latter:
\begin{equation}
 B_0(0,m^2,m^2)=\Delta_m\, .
\end{equation}
Note that in all the $f_i(q^2)$ form factors, the $B_0(i)$ functions appear in combinations such that their $1/\epsilon$ poles cancel each other, so these amplitudes are free of divergences.

\section{Bosonic form factors}
\label{BF}
The antisymmetric contribution to the photon propagator is characterized by the following form factors:
\begin{eqnarray}
g_1(q^2)&=&\frac{i\alpha}{4\pi}\Big\{B_0(1)+\frac{1}{4}B_0(2)+\frac{8m^2_W-3q^2}{3(4m^2_W-q^2)}\left[B_0(2)-B_0(1)\right] +\frac{16}{3}\left(1+\frac{q^2}{m^2_W}\right)m^4_WD_0(1)\nonumber \\
&&+\frac{8}{3}\left(1-\frac{q^2}{m^2_W}\right)m^4_WD_0(2)-\frac{4(2m^2_W-q^2)}{3(4m^2_W-q^2)}\Big\}\, ,
\end{eqnarray}
\begin{eqnarray}
g_2(q^2)&=&\frac{i\alpha}{4\pi}\Big\{B_0(1)+\frac{1}{4}B_0(2) +\frac{16m^2_W(4m^2_W-q^2)-15q^4}{15q^2(4m^2_W-q^2)}\left[B_0(2)-B_0(1)\right] \nonumber \\
 &&+\frac{8}{15}\left(12-\frac{4m^2_W}{q^2}+\frac{q^2}{m^2_W} \right)m^4_WD_0(1)-\frac{8}{15}\left(1+\frac{8m^2_W}{q^2}+\frac{3q^2}{m^2_W} \right)m^4_WD_0(2)\nonumber \\
 &&+\frac{4(16m^4_W+14m^2_Wq^2+3q^4)}{15q^2(4m^2_W-q^2)}\Big\}\, ,
\end{eqnarray}
\begin{eqnarray}
g_3(q^2)&=&\frac{i\alpha}{4\pi}\left[B_0(m^2_H)+B_0(m^2_Z) +\frac{1}{4}B_0(2) \right]\, ,\\
g_4(q^2)&=&\frac{i\alpha}{4\pi}\left(\frac{1}{4}B_0(2)\right)\, ,
\end{eqnarray}
\begin{eqnarray}
g_5(q^2)&=&\frac{i\alpha}{4\pi}\Big\{ -\frac{1}{6}\left[B_0(1)+\frac{1}{4}B_0(2)\right] +\frac{144m^4_W-136m^2_Wq^2+15q^4}{90q^2(4m^2_W-q^2)}\left[B_0(2)-B_0(1)\right]\nonumber \\
&&-\frac{1}{45}\left(37+\frac{36m^2_W}{q^2} +\frac{7q^2}{2m^2_W}\right)m^4_WD_0(1) -\frac{1}{45}\left(14+\frac{72m^2_W}{q^2} -\frac{23q^2}{m^2_W}\right)m^4_WD_0(2)\nonumber \\
&&+\frac{144m^4_W+136m^2_Wq^2-23q^4}{90q^2(4m^2_W-q^2)}\Big\}\, ,
\end{eqnarray}

\begin{eqnarray}
g_6(q^2)&=&\frac{i\alpha}{4\pi}\Big\{\frac{4}{15}\left(1-\frac{3m^2_W}{q^2}\right)-\frac{2m^2_W}{15q^2}\left[B_0(2)-B_0(1)\right]\nonumber \\
&&-\frac{1}{15}\left(2+\frac{q^2}{m^2_W}-\frac{24m^2_W}{q^2}\right)m^4_WD_0(1) -\frac{4}{15}\left(11-\frac{2q^2}{m^2_W}-\frac{12m^2_W}{q^2}\right)m^4_WD_0(2) \Big\}\, .
\end{eqnarray}

The form factors associated with the symmetric contribution obtained with the SM NLG are given by:
\begin{eqnarray}
\hat{g}_1(q^2)&=&\frac{i\alpha}{4\pi}\left\{\frac{1}{9}+\frac{1}{6}B_0(2)+\frac{4m^2_W}{3q^2}\left[B_0(2)-B_0(1)\right] \right\}\, ,\\
\hat{g}_2(q^2)&=&\frac{i\alpha}{4\pi}\left\{-\frac{2}{3}B_0(2)-\frac{4}{9}-\frac{32m^2_W}{4m^2_W-q^2}+
\frac{8m^2_W}{3q^2}\left(\frac{4m^2_W+5q^2}{4m^2_W-q^2}\right) \left[B_0(2)-B_0(1)\right]\right\}\, ,\\
\hat{g}_3(q^2)&=&\frac{i\alpha}{4\pi}\left\{\frac{1}{3}-\frac{2m^2_W}{q^2}\left[B_0(2)-B_0(1)\right] \right\} \, .
\end{eqnarray}

On the other hand, the form factors associated with the symmetric contribution obtained with the SME NLG are given by:

\begin{eqnarray}
\bar{g}_1(q^2)&=&\frac{i\alpha}{4\pi}\left\{\frac{B_0(2)}{6}+\frac{1}{9}+\frac{8m^2_W}{q^2}\left[B_0(1)-B_0(2)\right] \right\}\, , \\
\bar{g}_2(q^2)&=&\frac{i\alpha}{4\pi}\left\{-\frac{2B_0(2)}{3}-\frac{4}{9} -\frac{32m^2_W}{4m^2_W-q^2}+\frac{8m^2_W(4m^2_W-7q^2)}{3q^2(4m^2_W-q^2)}\left[B_0(1)-B_0(2)\right]\right\}\, , \\
\bar{g}_3(q^2)&=&\hat{g}_3(q^2)\, .
\end{eqnarray}

Finally, the form factors associated with gauge sector extension are given by:
\begin{eqnarray}
\tilde{g}_1(q^2)&=&\frac{i\alpha}{4\pi}\left\{-24B_0(2)-\frac{8q^2}{4m^2_W-q^2}-\frac{16m^2_W}{4m^2_W-q^2}\left[B_0(1)-B_0(2)\right]\right\}\, , \\
\tilde{g}_2(q^2)&=&\frac{i\alpha}{4\pi}\left\{-\frac{100}{3}B_0(2)-\frac{8}{9}-\frac{8q^2}{4m^2_W-q^2}
-\frac{16m^2_W}{3q^2}\left(\frac{4m^2_W+2q^2}{4m^2_W-q^2}\right)\left[B_0(1)-B_0(2)\right] \right\}\, ,\\
\tilde{g}_3(q^2)&=&\frac{i\alpha}{4\pi}\left\{\frac{13}{3}B_0(2)+\frac{2}{9}+\frac{2q^2}{3(4m^2_W-q^2)}+
\frac{4m^2_W}{3q^2}\left(\frac{4m^2_W+q^2}{4m^2_W-q^2}\right)\left[B_0(1)-B_0(2)\right]  \right\}\, ,\\
\tilde{g}_4(q^2)&=&\frac{i\alpha}{4\pi}\left\{\frac{2}{3}+\frac{4m^2_W}{q^2}\left[B_0(1)-B_0(2)\right]  \right\}\, .
\end{eqnarray}

The above expressions are given in terms of the following two- and four-point Passarino-Veltman scalar functions:
\begin{eqnarray}
B_0(1)&=&B_0(0,m^2_W,m^2_W) \, , \\
B_0(2)&=&B_0(q^2,m^2_W,m^2_W) \, , \\
B_0(m^2_H)&=&B_0(q^2,m^2_H,m^2_H) \, ,\\
B_0(m^2_Z)&=&B_0(q^2,m^2_Z,m^2_Z) \, ,
\end{eqnarray}
\begin{eqnarray}
D_0(1)&=&D_0(0,q^2,0,q^2,q^2,q^2,m^2_W, m^2_W,m^2_W,m^2_W)\, ,\\
D_0(2)&=&D_0(0,0,q^2,q^2,0,q^2,m^2_W, m^2_W,m^2_W,m^2_W)\, ,
\end{eqnarray}
where the four-point scalar functions is defined as~\cite{FeynCalc}:
\begin{equation}
D_0(p^2_1,p^2_{12},p^2_{23},p^2_3,p^2_2,p^2_{13},m^2_0,m^2_1,m^2_2,m^2_3)=\int \frac{d^Dk}{i\pi} \frac{1}{[k^2-m^2_0][(k+p_1)^2-m^2_1][(k+p_2)^2-m^2_2][(k+p_3)^2-m^2_3]} \, , \nonumber
\end{equation}
with $p^2_{ij}=(p_i-p_j)^2$.

\end{document}